\documentclass[journal]{IEEEtran}
\usepackage{amsmath,amsfonts}
\usepackage{algorithm}
\usepackage{algorithmicx}
\usepackage{algpseudocode}
\usepackage{array}
\usepackage{textcomp}
\usepackage{stfloats}
\usepackage{url}
\usepackage{verbatim}
\usepackage{graphicx}
\usepackage{cite}
\usepackage{booktabs}
\usepackage{xcolor}
\usepackage{makecell}
\usepackage{multirow}

\usepackage[table]{xcolor}
\usepackage{threeparttable}
\newcolumntype{L}[1]{>{\raggedright\arraybackslash}p{#1}} 
\newcolumntype{C}[1]{>{\centering\arraybackslash}p{#1}}   
\newcolumntype{R}[1]{>{\raggedleft\arraybackslash}p{#1}}

\makeatletter
\let\MYcaption\@makecaption
\makeatother

\usepackage[font=footnotesize]{subcaption}

\makeatletter
\let\@makecaption\MYcaption
\makeatother

\begin{document}

\title{Adaptive End-to-End Transceiver Design for NextG \\ Pilot-Free and CP-Free Wireless Systems}

\author{Jiaming Cheng,~\IEEEmembership{Student Member,~IEEE}, Wei Chen,~\IEEEmembership{Senior Member,~IEEE}, and Bo Ai,~\IEEEmembership{Fellow,~IEEE}

\thanks{Jiaming Cheng, Wei Chen and Bo Ai are with State Key Laboratory of Advanced Rail Autonomous Operation, and the School of Electronic and Information Engineering, Beijing Jiaotong University, China. (e-mail:
 \{jiamingcheng, weich, boai\}@bjtu.edu.cn)}
 
 }

\maketitle

\begin{abstract}
The advent of artificial intelligence (AI)-native wireless communication is fundamentally reshaping the design paradigm of next-generation (NextG) systems, where intelligent air interfaces are expected to operate adaptively and efficiently in highly dynamic environments. Conventional orthogonal frequency division multiplexing (OFDM) systems rely heavily on pilots and the cyclic prefix (CP), resulting in significant overhead and reduced spectral efficiency. To address these limitations, we propose an adaptive end-to-end (E2E) transceiver architecture tailored for pilot-free and CP-free wireless systems. The architecture combines AI-driven constellation shaping and a neural receiver through joint training. To enhance robustness against mismatched or time-varying channel conditions, we introduce a lightweight channel adapter (CA) module, which enables rapid adaptation with minimal computational overhead by updating only the CA parameters. Additionally, we present a framework that is scalable to multiple modulation orders within a unified model, significantly reducing model storage requirements. Moreover, to tackle the high peak-to-average power ratio (PAPR) inherent to OFDM, we incorporate constrained E2E training, achieving compliance with PAPR targets without additional transmission overhead. Extensive simulations demonstrate that the proposed framework delivers superior bit error rate (BER), throughput, and resilience across diverse channel scenarios, highlighting its potential for AI-native NextG.
\end{abstract}


\section{Introduction}
\IEEEPARstart{W}{ith} the rapid evolution of wireless communication technologies, the demand for higher spectral efficiency, lower latency, and improved robustness continues to grow~\cite{chen2024signal,wang2023road}. In response to these highly anticipated requirements, the next-generation (NextG) (e.g., the sixth generation (6G) and beyond) networks are envisioned to integrate advanced technologies such as artificial intelligence (AI) and machine learning (ML)~\cite{chen20235g}, enabling more adaptive, intelligent, and efficient communication systems~\cite{gunduz2019machine}. AI/ML technologies are driving a fundamental paradigm shift in wireless system design, evolving from auxiliary tools into native design elements~\cite{hoydis2021toward}. This transformation extends beyond local performance optimization, promoting an end-to-end (E2E) reconfiguration of the network architecture that embeds intelligence across the entire lifecycle of communication systems. It signifies the emergence of AI-native air interface design as a cornerstone of NextG wireless communications~\cite{hoydis2021toward}.

As a key initiative, the 3rd generation partnership project (3GPP) has launched dedicated efforts to explore AI/ML integration into the radio access network (RAN), aiming to enhance system performance, reduce complexity, and improve scalability~\cite{chen20235g}. 3GPP has initiated research into AI/ML-enhanced technologies across specific use cases, such as channel state information (CSI) feedback enhancements~\cite{guo2024deep,3gpp_bjtu}, beam management, and positioning accuracy enhancements. All these use cases exhibit significant potential and are well-suited for integration with AI~\cite{xu2022deep,cheng2023swin}.

In the domain of physical layer transceiver design, conventional transceivers adopt a modular structure to ensure operational stability, but this design often results in inter-module dependencies, poor adaptability, and less-than-optimal performance. Recently, research has increasingly turned to AI-driven architectures that aim to break these modular barriers. A novel E2E learning paradigm has been proposed in~\cite{o2017introduction}, enabling joint optimization of transmitter and receiver tailored to specific channel environments. The concept of neural receivers, where a single neural network is trained to jointly perform channel estimation, equalization, and demapping, is introduced in~\cite{honkala2021deeprx} and demonstrates superior performance compared to traditional receivers. By embedding neural networks into the signal processing chain in a principled and integrated manner, these approaches aim to overcome fundamental limitations of traditional model-based methods.

\subsection{Motivation}
In the fifth-generation (5G) system, pilot signals are essential for ensuring reliable and effective communication. For instance, demodulation reference signals (DMRS) are used to enable accurate channel estimation. These pilots are predefined sequences that are orthogonally allocated with data in the time-frequency resource grid. This arrangement leads to resource contention and significant overhead, thereby reducing spectral efficiency and limiting system throughput. With the advent of NextG networks, featuring massive multiple-input multiple-output (MIMO) configurations, ultra-high mobility, and more complex wireless environments~\cite{wang2023road}, the pressure on pilot design and overhead will become even more pronounced. This intensifies the resource contention between pilots and data transmission. Additionally, in conventional orthogonal frequency division multiplexing (OFDM) systems, a cyclic prefix (CP) is inserted to mitigate inter-symbol interference (ISI), but it further degrades spectral efficiency due to the inclusion of redundant data. 

Focusing on the inefficiencies caused by pilot overhead and CP redundancy, it becomes imperative to move beyond conventional transmission designs and explore AI-native strategies for more efficient and adaptive communication. In this context, transmission schemes with superimposed pilots have been proposed to enhance the system throughput~\cite{ma2017orthogonal, aoudia2021end, han2025interference}. This architecture suffers from interference between pilot and data signals, which limits the overall system performance. Meanwhile, determining the optimal power allocation between pilot and data symbols further increases the system design complexity.
In addition, an E2E transceiver architecture proposed in~\cite{aoudia2021end} integrates an autoencoder-based neural network with a learnable constellation for OFDM systems. This approach achieves state-of-the-art performance over realistic wireless channels without requiring pilots. An E2E solution for frequency-selective channels is proposed in~\cite{ye2021deep}, which bypasses the use of pilots for channel estimation. Furthermore, to enhance spectral efficiency, CP is omitted in~\cite{ye2017power}, with pilots still employed. Extending these ideas, the removal of both CP and pilots is addressed simultaneously in~\cite{ait2021trimming}, demonstrating that E2E learning enables the elimination of these overhead components and leads to significant throughput improvements. However, these studies do not consider issues such as adaptive re-training and online learning, modulation-order switching, or practical hardware constraints on transmission power.

Recent standardization activities have also demonstrated increasing interest in AI/ML-based solutions for the air interface design. In particular, the 3GPP community has initiated extensive discussions on AI-native air interface in Release 20 recently.
These efforts encompass several promising use cases, including E2E learning with autoencoders~\cite{3gpp_nvidia}, overlaid DMRS and data transmission schemes~\cite{3gpp_huawei}, and pilot-free AI-enabled approaches for joint modulation and equalization~\cite{3gpp_deepsig}. These directions highlight the potential of AI/ML techniques in redefining air interface design. However, there are still critical challenges in realizing adaptive E2E transceivers for pilot- and CP-free systems in practical deployment. These include coping with dynamic channel conditions, achieving scalability across modulation orders, and maintaining a low peak-to-average power ratio (PAPR).

\subsection{Challenges and Related Works}
Traditional AI models are typically trained for specific scenarios and require large amounts of data to generalize effectively~\cite{xu2024learning}. Transfer learning offers a promising solution by leveraging knowledge from existing data and models to adapt learned representations to new communication environments~\cite{ahmadi2025towards,uyoata2024transfer}. This scheme enables improved generalization with fewer data samples and reduced training effort~\cite{nguyen2022transfer}. However, current E2E approaches to signal transmission rely on computation-intensive full fine-tuning, resulting in high computational cost and increased risk of overfitting during transfer learning, particularly when the target domain has limited data~\cite{yang2023aim}. These limitations become particularly critical in practical deployments, where only limited channel data is available for adaptation~\cite{wang2023few,you2021towards}.

In practical systems, different modulation orders lead to changes in the input and output dimensions in the model. Training a single model for each modulation order incurs large computational and storage overhead, hindering deployment and maintenance. While a scalable modulation order mechanism is proposed for the receiver side in~\cite{honkala2021deeprx,han2025interference}, the transmitter still relies on conventional modulation schemes without considering learnable constellations. To meet practical requirements, it is essential to design a flexible E2E transceiver solution capable of handling various modulation orders within a single unified network.

Moreover, the high PAPR will induce nonlinear distortion in hardware and lead to inefficiencies in power utilization, which is especially problematic in energy-constrained applications such as mobile and Internet of Things (IoT) devices~\cite{psomas2024wireless}. As a result, addressing the PAPR issue is particularly important in uplink transmissions. Traditional PAPR reduction techniques are generally categorized into distortion-based and distortionless methods. The signal distortion techniques, including clipping and filtering~\cite{gutman2013iterative}, limit the time-domain peak envelope to a specified threshold. The signal non-distortion techniques, including selective mapping and partial transmit sequence~\cite{baxley2007comparing}, require side information (SI) to recover the original signal, introducing additional bandwidth overhead. Meanwhile, errors in SI detection can severely degrade the bit error rate (BER) performance. Recent studies have integrated deep learning (DL) into waveform design for effective PAPR reduction. In~\cite{ait2022waveform, marasinghe2024constellation}, the authors apply constellation shaping to single-carrier waveforms over multipath channels for joint PAPR reduction and achievable rate maximization. An E2E convolutional-autoencoder learning model is proposed in~\cite{huleihel2024low}, which utilizes a single PAPR reduction block. While prior work has extensively explored PAPR reduction in conventional OFDM systems, there remains a lack of effective solutions tailored to pilot-free and CP-free systems.

\subsection{Contributions}
Building on these advancements, this paper introduces an adaptive E2E transceiver for pilot-free and CP-free OFDM systems. By integrating AI-based constellation shaping with receiver design, the transceiver enables joint optimization of key components such as mapper, channel estimation, equalization, and demapper, leading to improved overall performance. The proposed transceiver incorporates multiple innovative mechanisms to address the challenges of practical deployments. The contributions of this paper are summarized as follows:

\begin{itemize} 
\item \textit{Parameter-Efficient Adaptation for Dynamic Environments}: We propose a lightweight, plug-and-play channel adapter integrated into the receiver design that enables efficient adaptation to highly dynamic environments by fine-tuning only a few parameters. When transferring to a new environment, the channel adapter learns site-specific feature modulations on the intermediate representations of backbones while keeping the pre-trained parameters frozen. The proposed adapter can also incorporate auxiliary information, such as noise power, to further enhance adaptation efficiency and noise robustness.
\item \textit{Storage-Efficient Adaptation for Multi-Order Modulation}: We develop a scalable mechanism for geometric constellation shaping and receiver design across multiple modulation orders. This allows a single unified model to operate effectively under various modulation schemes, significantly reducing model storage overhead and simplifying model lifecycle management.
\item \textit{PAPR-Constrained E2E Learning}: We also investigate waveform optimization and reliable transmission in pilot-free and CP-free systems under PAPR constraints. To address this issue, we utilize learning-based geometric shaping to design a power-efficient transmit waveform, enabling low-complexity implementation at the transmitter without relying on deep neural networks. The resulting E2E system achieves PAPR reduction and competitive BER performance compared to the conventional schemes. 
\item \textit{Performance Validation}: We carry out extensive simulations on 3GPP-compliant channel models with different pilot and CP configurations. We compare the BER and throughput performance of different schemes under various user mobility speeds and consistently observe that the proposed adaptive E2E transceiver achieves significant performance gains, which may hopefully provide valuable insights for future standardization efforts.
\end{itemize}

The rest of this paper is organized as follows. Section \uppercase\expandafter{\romannumeral2} introduces the baseline and AI-based transceiver architectures, with particular attention to the PAPR problem in the OFDM system. Section \uppercase\expandafter{\romannumeral3} introduces the proposed adaptive AI-based transceiver network and the training methodology, while our experimental results are discussed in Section \uppercase\expandafter{\romannumeral4}. Finally, Section \uppercase\expandafter{\romannumeral5} concludes this paper.

\section{System Model}

\begin{figure*}[!t]
\centering
\includegraphics[width=0.98\textwidth]{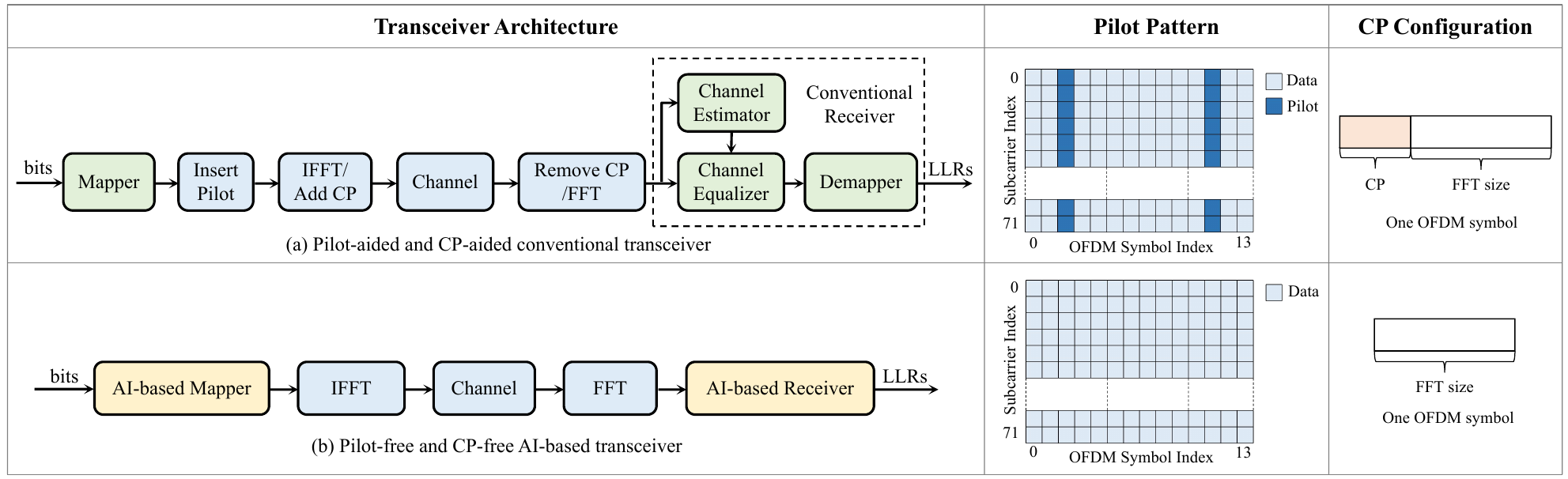}
\caption{Overview of conventional (pilot-aided and CP-aided) and AI-based (pilot-free and CP-free) transceiver architectures.}
\label{Fig: architecture}
\end{figure*}

We consider a typical uplink single-input multiple-output (SIMO) system with a single transmitting antenna at the user equipment (UE) and $N_r$ receiving antennas at the base station (BS), operating in a single stream configuration. $N_c$ subcarriers with $N_s$ consecutive OFDM symbols are allocated. In this section, the conventional and AI-based transceivers are introduced. Apart from introducing the neural network-based receiver, the AI-based transceiver can be directly integrated into 5G~NR systems simply through customized constellations as well as pilot-free and CP-free configurations.
\subsection{Baseline Transceiver}
Conventional systems usually adopt a transceiver architecture that relies on pilots and CP, as shown in Fig.~\ref{Fig: architecture}(a). The transmission bits are first modulated using quadrature amplitude modulation (QAM) with a modulation order of $2^M$, where $M$ denotes the number of bits per symbol. After modulation and pilot insertion, the symbol undergoes an inverse fast Fourier transform (IFFT), followed by the addition of the CP to mitigate ISI and intercarrier interference (ICI). The resulting signal is then transmitted through the channel. At the receiver side, the CP is removed, and a fast Fourier transform (FFT) is applied to recover the signal.

Under this framework, the received signal at the $i$-th OFDM symbol and the $j$-th subcarrier can be expressed as
\begin{equation} \mathbf{y}_{ij} = \mathbf{h}_{ij} x_{ij} + \mathbf{n}_{ij}, \label{eq:pilot_rb_estimate} \end{equation}
where $\mathbf{y}_{ij}, \mathbf{h}_{ij}, \mathbf{n}_{ij} \in \mathbb{C}^{N_r}$ are the received signals, the channel coefficients, and the additive white Gaussian noise with variance of $N_0=\sigma^2$, respectively. 
The transmitted signal is represented by  $x_{ij} \in \mathbb{C}$.

The UE transmits pilot symbols over designated subcarriers and time slots, and the index set of pilot positions can be denoted as $\mathcal{P}$. The least squares (LS) estimate of the channel at pilot positions is then computed as
\begin{equation} \hat{\mathbf{h}}_{ij} = \frac{\mathbf{y}_{ij}}{x_{ij}}, \quad (i,j) \in \mathcal{P}. \end{equation}
Since pilots are sparsely distributed, the full channel matrix is reconstructed over the OFDM grid using linear interpolation. Once the channel is estimated, linear minimum mean square error (LMMSE) equalization is applied at the BS to suppress the effects of fading and noise. For each OFDM symbol $i$ and each subcarrier $j$, the equalized symbol grid is obtained as
\begin{equation} \hat{x}_{ij} = \left(\hat{\mathbf{h}}_{i,j}^H \hat{\mathbf{h}}_{i,j} + \frac{\sigma^2}{E_s}\right)^{-1} \hat{\mathbf{h}}_{i,j}^H \mathbf{y}_{ij}, \end{equation}
where $\hat{x}_{ij}$ denotes the equalized signal. The recovered symbols are soft-demapped into log-likelihood ratios (LLRs) under the Gaussian noise assumption, which are then passed to the channel decoder to recover the transmitted bits.

\subsection{AI-based Constellation Shaping}
In contrast to conventional systems with pilot and CP overhead, we consider a pilot-free and CP-free AI-based transceiver architecture~\cite{ait2021trimming}, as illustrated in Fig.~\ref{Fig: architecture}(b).
Instead of using a fixed modulation constellation, the transmitter learns constellation points as trainable parameters through E2E training. Specifically, two real-valued vectors, $\mathbf{c}_\mathrm{Re} \in \mathbb{R}^{2^{M}}$ and $\mathbf{c}_\mathrm{Im} \in \mathbb{R}^{2^{M}}$, are jointly trained. To accelerate convergence and ensure a reasonable initial performance, these trainable constellation points are initialized using the standard QAM-constellation points. The resulting complex-valued constellation points are expressed as $\mathbf{c} = \mathbf{c}_\mathrm{Re} + j \mathbf{c}_\mathrm{Im}$, which are then normalized and centered following the procedure in~\cite{aoudia2021end}, which can be written as
\begin{equation} \bar{\mathbf{c}} = \frac{\mathbf{c} - \frac{1}{2^{M}}\sum_{i=1}^{2^{M}} \mathbf{c}_i}{\sqrt{\frac{1}{2^{M}}\sum_{i=1}^{2^{M}} \left|\mathbf{c}_i \right|^2 - \left|\frac{1}{2^{M}}\sum_{i=1}^{2^{M}} \mathbf{c}_i\right|^2}}. \end{equation}
Centering the constellation effectively mitigates potential direct current (DC) offset. Moreover, the learned constellations are normalized to unit energy, ensuring that learning-based geometric
shaping preserves the same total transmit energy as the conventional OFDM system. This AI-driven mapping strategy enables the transmitter to adapt the constellation geometry to channel conditions and the loss function. 

\subsection{AI-based Receiver}
At the receiving end, the neural receiver is employed after the FFT operation to replace the signal processing modules for channel estimation, equalization, and demapping. The input to the network is the received resource grid, denoted as $\mathbf{Y} \in \mathbb{C}^{N_r \times N_s \times N_c}$, and the noise information $\mathbf{N}_0\in \mathbb{R}^{N_s \times N_c}$. The output is a tensor of LLRs, represented by $\mathbf{L} \in \mathbb{R}^{M \times N_s \times N_c}$. A detailed description of the neural receiver architecture will be presented in Section~\ref{sec:adaptive_channel}.
By jointly training the constellation and the neural receiver, the system effectively compensates for the absence of pilots and CP.

\begin{figure*}[!t]
\centering
\includegraphics[width=0.95\textwidth]{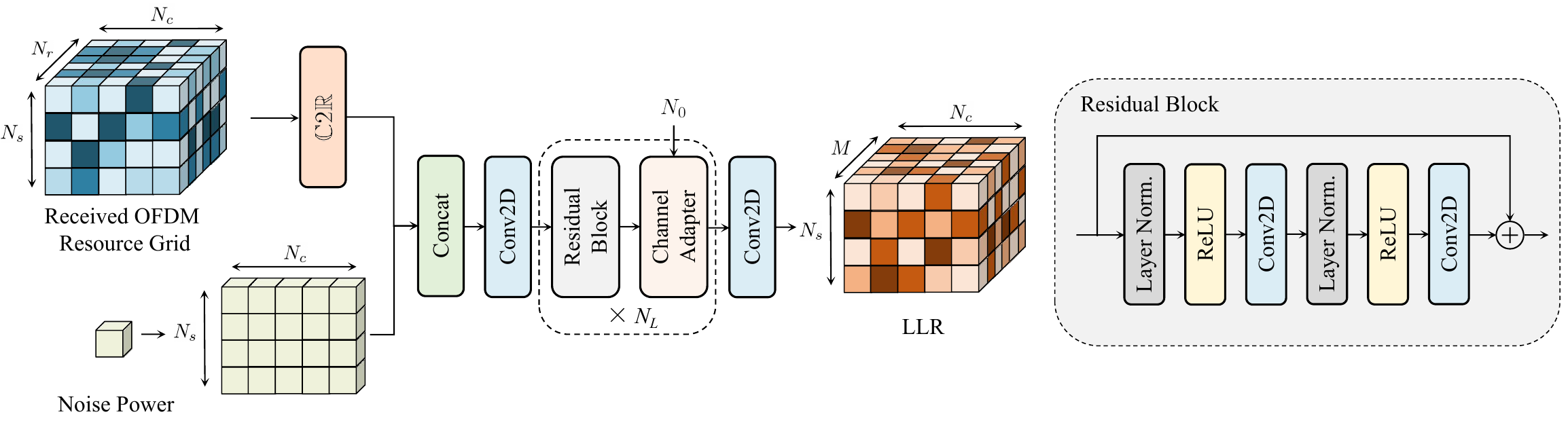}
\caption{Illustration of the proposed neural receiver architecture and the structure of the residual block.}
\label{Fig: neural receiver}
\end{figure*}

\subsection{PAPR in the OFDM System}
As demonstrated in~\cite{li2011novel,marasinghe2024constellation}, constellation shaping can be leveraged to reduce the PAPR. This insight motivates the incorporation of a PAPR constraint into our E2E system design.
In an OFDM system with $N_c$ subcarriers, the discrete-time OFDM signal is obtained via an IFFT, which is written as
\begin{equation}
    \tilde{x}_n = \frac{1}{\sqrt{{LN_c}}}\sum_{k=0}^{{LN_c}-1}{X}_{k}e^{j\frac{2\pi}{{L}{N_c}}kn}, \;\;  0\le n\le {L}{N_c}-1, \;\;\label{IFFT}
\end{equation}
where $X_k$ denotes the frequency-domain symbol. The factor ${L}\geq1$ represents the oversampling rate. 
$LN_c$-point oversampling is achieved by adding $(L-1)N_c$ zeros to the $N_c$-point frequency-domain signal and then applying the IFFT to the resulting $LN_c$-point sequence.

To reduce computational complexity and conserve resources, the signal transmission is performed at the Nyquist sampling rate without employing oversampling. However, to enable accurate PAPR evaluation, the transmitted signal is oversampled by a factor of $L=4$ during PAPR computation, as recommended in~\cite{jiang2008overview}. This oversampling provides a closer approximation to the continuous-time OFDM waveform, thereby yielding more reliable PAPR estimates. By decoupling the oversampling process from actual transmission, the system achieves efficient signal delivery while maintaining the fidelity of PAPR assessment.

The PAPR of the transmitted signal in \eqref{IFFT} is defined as the ratio of the maximum peak power to the average power of the OFDM signal, which can be expressed as
\begin{equation}
\text{PAPR} = \frac{\max_{0 \le n \le {LN_c}-1} |{\tilde{x}_n}|^2}{\mathbb{E}[|\tilde{x}_n|^2]},
\end{equation}
where the expectation is over the oversampled signal $\tilde{x}_n$.

\section{Adaptive AI Transceiver Design}
In this section, we describe the proposed adaptive E2E transceiver in detail, which includes an effective adaptation design for new environments and a scalable mechanism for supporting multiple modulation orders. Then, we introduce the loss function that incorporates PAPR constraints and illustrate the training framework of the proposed transceiver.

\subsection{Adaptive Channel Scenario}\label{sec:adaptive_channel}
We design a neural receiver capable of adapting to varying channel conditions, as shown in Fig.~\ref{Fig: neural receiver}. The proposed neural receiver processes a three-dimensional input tensor constructed from both the received resource grid $\mathbf{Y}$ and the noise information $\mathbf{N}_0$. The real and imaginary parts of the complex-valued resource elements (REs) are separated and concatenated along the channel dimension. Furthermore, the noise power $N_0$ is broadcast across the time and frequency dimensions to form a supplementary channel of size $N_s \times N_c$, and then appended as an auxiliary input, which enables a tunable balance between the information content of the conveyed features and their robustness to channel noise. Consequently, the final input tensor has dimensions of $(2N_r+1) \times N_s \times N_c$, comprising $2N_r$ channels from the signal components and one additional channel for the noise information.

In our proposed architecture, we adopt $N_L$ Residual blocks as the backbone, which has been demonstrated to be effective in other works~\cite{aoudia2021end},~\cite{cammerer2023neural}. As depicted in Fig.~\ref{Fig: neural receiver}, each block consists of double sequential layer normalizations, ReLU activations, and two-dimensional convolutional layers (Conv2D) with residual connections in each block~\cite{he2016deep}. The choice of convolutional neural networks (CNNs) is motivated by their natural suitability for OFDM waveforms. OFDM signals can be represented in 2D space along the subcarrier and OFDM symbol axes, making CNNs ideal for learning translationally invariant operations. 
The residual CNN backbone is adopted to stabilize training and better capture the complex time–frequency structures in wireless channels. 
Each Residual block is followed by a lightweight channel adapter module, which will be presented later in this subsection. The final layer outputs bit-wise LLRs via a Conv2D, corresponding to the current modulation.

\begin{figure}[!t]
\centering
\includegraphics[width=0.48\textwidth]{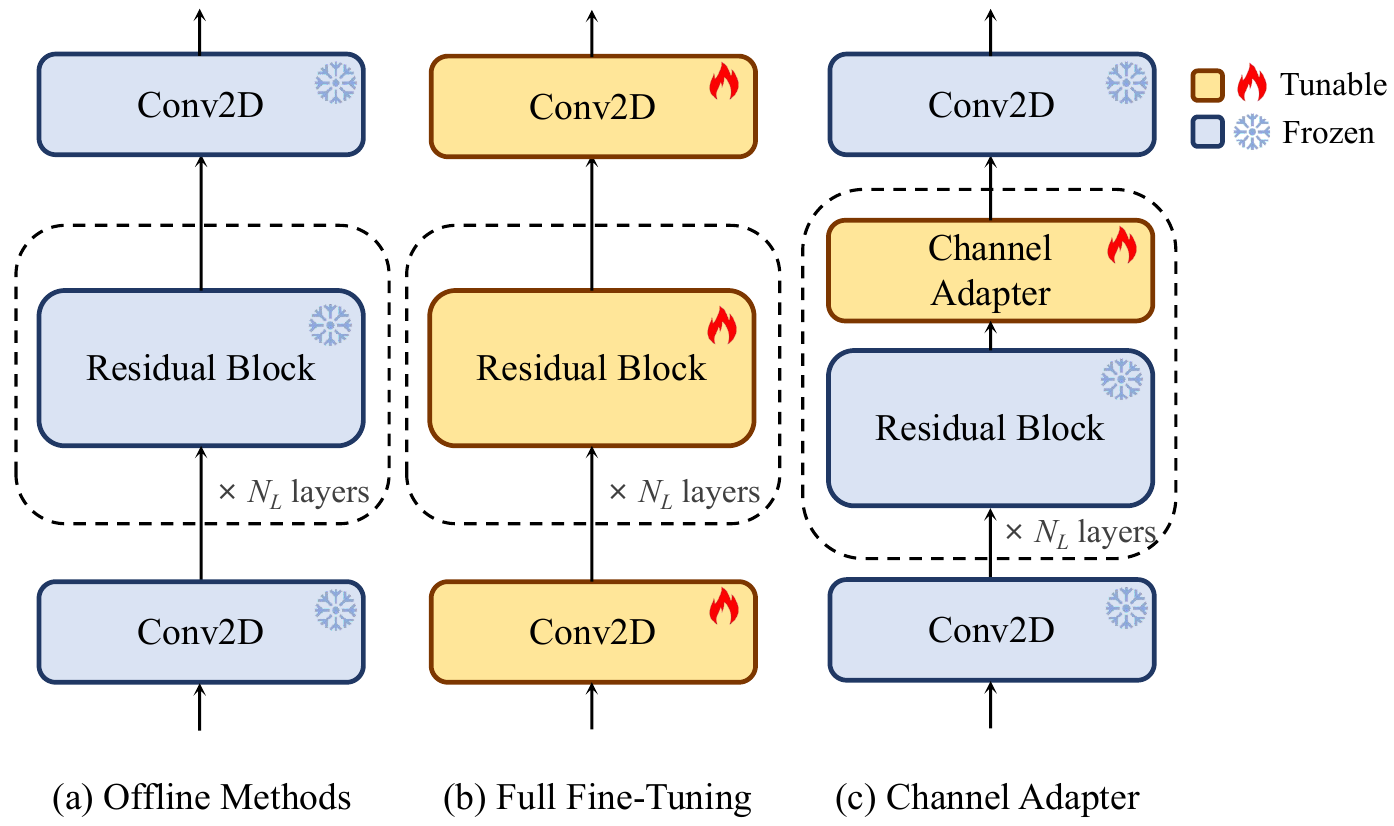}
\caption{Comparative illustration of our proposed channel adapter framework versus popular and widely used alternatives. (a) depicts the typical offline method. (b) illustrates the traditional end-to-end approach with full fine-tuning. (c) Tailored for the symbol detection task, our network incorporates a lightweight adapter within the backbone to enable efficient transfer learning.}
\label{Fig: adapter}
\end{figure}

In practical deployments, it is critical for E2E neural communication systems to adapt to dynamic channel conditions with limited computational resources and only a small number of observed channel samples~\cite{3gpp_nvidia_ran108}.
Fig.~\ref{Fig: adapter} presents a comparative overview of representative transfer learning strategies in the context of E2E learning. The conventional offline scheme, illustrated in Fig.~\ref{Fig: adapter}(a), relies on pretraining the model under fixed channel conditions and deploying it without further updates. An alternative is the full fine-tuning strategy, illustrated in Fig.~\ref{Fig: adapter}(b), where the entire network is updated using measured data. 
Although this approach offers strong adaptability, it requires extensive training data and incurs substantial computational overhead for each new channel condition, which limits its practicality for real-time adaptation.
Moreover, as noted in~\cite{yang2023aim}, full fine-tuning may cause overfitting or catastrophic forgetting, especially for large pretrained models, and can degrade performance when the available channel samples lack sufficient diversity.
To address these limitations, we propose fine-tuning a lightweight, plug-and-play module named channel adapter (CA), as depicted in Fig.~\ref {Fig: adapter}(c). By updating only a small subset of parameters, the CA module enables efficient and effective transfer learning, striking a favorable balance between adaptability and resource efficiency.

\begin{figure}[!t]
\centering
\includegraphics[width=0.26\textwidth]{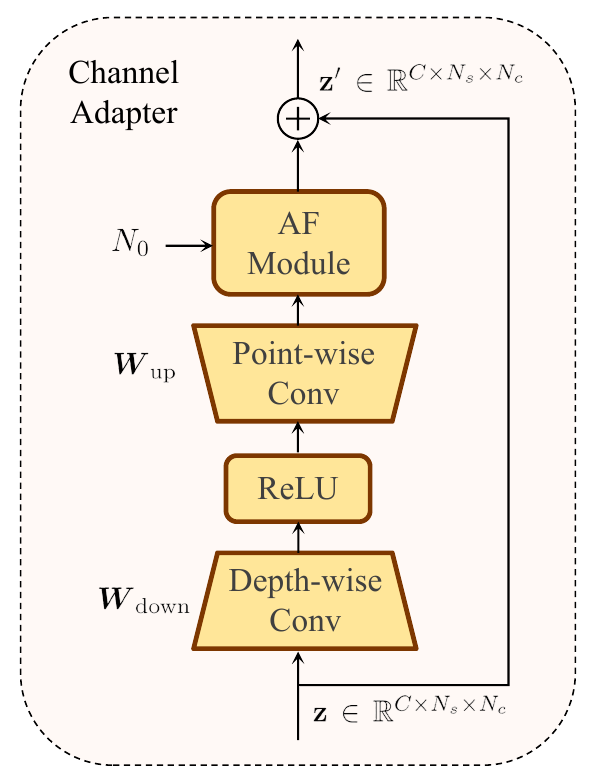}
\caption{Architecture of channel adapter, which employs a bottleneck structure composed of depth-wise separable convolutions and ReLU activation, followed by the AF module.}
\label{Fig: channel adapter}
\end{figure}

The architecture of CA follows the general bottleneck design of the standard adapter~\cite{houlsby2019parameter}. As illustrated in Fig.~\ref {Fig: channel adapter}, the architecture consists of two convolutional layers with a ReLU activation function applied in between, followed by an attention feature (AF) module proposed in~\cite{xu2021wireless}. The first convolution performs channel dimension reduction, while the second convolution restores the original channel dimension. To further reduce parameter overhead, we employ depthwise separable convolutions~\cite{howard2017mobilenets} within the Channel Adapter. Specifically, the first layer uses a depthwise convolution with weights $\mathbf{W}_{\text{down}} \in \mathbb{R}^{\frac{C}{\gamma} \times \gamma \times K \times K}$, and the second layer uses a pointwise convolution with weights $\mathbf{W}_{\text{up}} \in \mathbb{R}^{C \times \frac{C}{\gamma} \times 1 \times 1}$, where $\gamma$ denotes the channel reduction ratio, $K$ is the kernel size, and $C$ represents the channel dimension, identical for both input and output. The non-linear activation function $\boldsymbol{\sigma}$ is inserted between these two convolutional layers. Furthermore, the AF module is integrated to mitigate performance degradation under varying noise levels, and it generates noise-aware weight $\boldsymbol{\alpha} \in \mathbb{R}^{C}$, which is applied to the input features via channel-wise multiplication. Afterward, a residual connection is added to the output of the AF module. Note that $\mathbf{z}$ and $\mathbf{z}^{\prime}$ are the input and output features with the same shape $\mathbb{R}^{C\times N_s \times N_c}$. The overall computation of the adapter module can be formulated as
\begin{align}
\hat{\mathbf{z}} &= \boldsymbol{\sigma}(\mathbf{W}_{\text{down}} \hat{\otimes} \mathbf{z}), \\
\tilde{\mathbf{z}} &= \mathbf{W}_{\text{up}} \dot{\otimes} \hat{\mathbf{z}}, \\
\mathbf{z}^{\prime} &= \boldsymbol{\alpha}\cdot\tilde{\mathbf{z}} + \mathbf{z},
\end{align}
where $\dot{\otimes}$ and $\hat{\otimes}$ denotes point-wise and depth-wise convolution, respectively. The learnable weight $\boldsymbol{\alpha}$ is calculated according to the output feature of point-wise convolution and the noise power $N_0$~\cite{xu2021wireless}.
This design ensures robust performance across varying channel conditions while maintaining practicality for real-world deployment.

\subsection{Adaptive Multi-Order Modulation}
To support multiple modulation orders in practical systems, we propose a unified AI transceiver architecture, which significantly reduces network storage overhead. The AI transceiver is designed based on the maximum modulation order $M_{\text{max}}$, while the actual modulation order $M$ is provided as an auxiliary input to enable dynamic adaptation. The selection of modulation order is determined based on the corresponding block error rate (BLER) and throughput under different channel conditions.

\begin{figure}[!t]
\centering
\begin{subfigure}[b]{0.48\textwidth}
    \centering
    \includegraphics[width=0.86\textwidth]{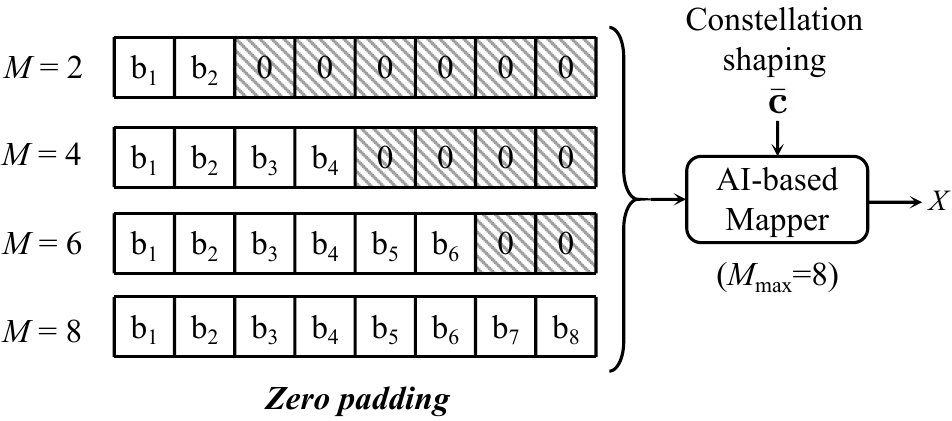}
    \caption{Geometric
constellation with zero padding at the transmitter}
    \label{Fig: multiorder_aitx}
\end{subfigure}

\vspace{0.08cm} 

\begin{subfigure}[b]{0.48\textwidth}
    \centering
    \includegraphics[width=0.85\textwidth]{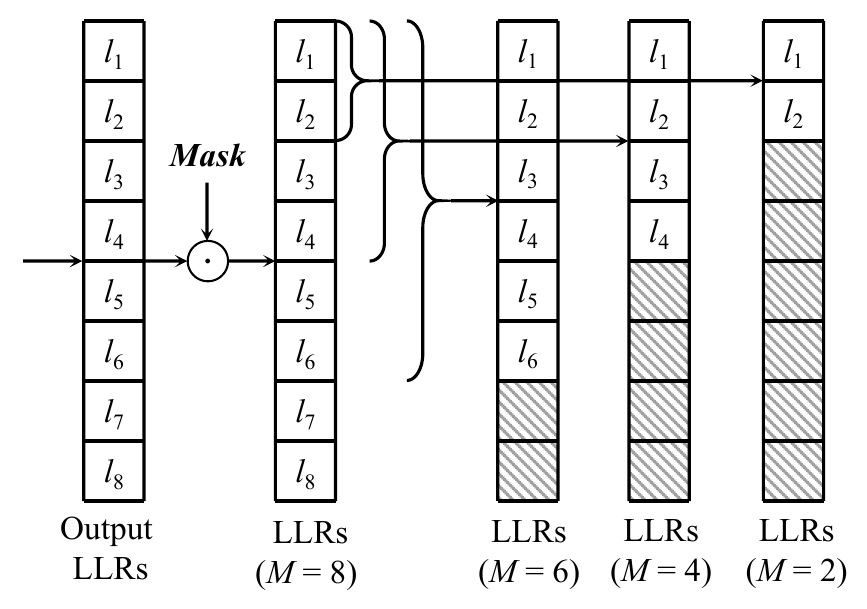}
    \caption{Masking LLRs at the receiver}
    \label{Fig: multiorder_airx}
\end{subfigure}

\caption{AI transceiver design for supporting multiple modulation orders: (a) transmitter-side geometric constellation shaping using zero padding; (b) receiver-side LLR masking.}
\label{Fig: multiorder architecture}
\end{figure}

To facilitate constellation mapping across varying modulation orders, the input bitstream at the transmitter is reshaped into groups of \( M \) bits for each time-frequency grid \( (i,j) \), and the resulting data is represented as a tensor \( \mathbf{B} \in \{0,1\}^{M \times N_s \times N_c} \), where \( M \) denotes the number of bits per symbol.  For modulation orders \( M < M_{\text{max}} \), each bit group is zero-padded to length \( M_{\text{max}} \) to allow consistent indexing over a shared non-uniform custom constellation set \( \bar{\mathbf{c}} \in \mathbb{C}^{2^{M_{\text{max}}}} \), as illustrated in Fig.~\ref{Fig: multiorder architecture}(\subref{Fig: multiorder_aitx}). 

Each zero-padded bit group is then interpreted as an $M_{\text{max}}$-bit binary number and converted into an integer index \( I_{ij} \in \left\{ 0, 2^{M_{\text{max}}-M}, \dots, \left(2^M-1\right)\cdot 2^{M_{\text{max}}-M} \right\} \). The effective constellation mapping is thus given by:
\begin{equation}
    X_{ij} = \bar{\mathbf{c}}[I_{ij}].
\end{equation}
To ensure only \( 2^M \) valid constellation points are used, we construct a modulation-order-specific subset $\mathcal{C}_M$  by uniformly sampling from \( \bar{\mathbf{c}} \) with a step size \( 2^{M_{\text{max}} - M} \). Note that the power constraint is imposed on the full constellation set corresponding to the maximum modulation order, ensuring that the resulting constellation maintains unit average power. For cases where the modulation order $M<M_{max}$, the constellation subset $\mathcal{C}_M$ may not be strictly power-normalized. Let the average symbol power over $\mathcal{C}_M$ be denoted as
\begin{equation}
    \mathbb{E}_{c_i \in \mathcal{C}_M} \left[ |c_i|^2 \right] = P_0^{(M)}.\;\;  M<M_{max}
\end{equation}
Then the noise for the modulation order $M$ should be adjusted accordingly:
\begin{equation}
\tilde{n} = \sqrt{P_{0}^{(M)}} \cdot n,
\end{equation}
where $n \sim \mathcal{CN}(0, \sigma^2)$ is the original complex Gaussian noise and $\tilde{n}$ is the scaled noise that matches the effective signal power $P_0$.
This design enables seamless modulation order adaptation by allowing the neural transmitter to learn unified constellation mappings, facilitating integration with link adaptation mechanisms informed by channel state or higher-layer scheduling. Moreover, it ensures a fair performance comparison across different modulation orders, since the mapping is learned under a shared training framework and power constraint.

To enable flexible adaptation to various modulation schemes within a unified receiver architecture, the neural network is designed to output a redundant bit-wise LLR tensor denoted as $\mathbf{Z} \in \mathbb{R}^{M_{\text{max}} \times N_s \times N_c}$. A modulation-aware mask dynamically selects the relevant $M$ bit positions based on the current modulation order, as illustrated in Fig.~\ref{Fig: multiorder architecture}(\subref{Fig: multiorder_airx}).

The masking operation employs a learnable weight tensor $\mathbf{W} \in \mathbb{R}^{{M_{\text{max}}} \times N_s \times N_c}$, which is normalized through a sigmoid activation to produce a soft mask $\mathbf{W}^* = \operatorname{Sigmoid}(\mathbf{W})$. The masking mechanism enables a single neural architecture to operate across multiple modulation schemes without architectural modifications, while adapting to the asymmetric geometric coordinates of constellation points learned at the transmitter. The final LLR output $\mathbf{L} \in \mathbb{R}^{M \times N_s \times N_c}$ is then obtained by
\begin{equation}
\mathbf{L}_m=\mathbf{Z}_m \circ \mathbf{W}^{*}_m, \;\;  0\le m\le M,
\end{equation}
where $\circ$ is the Hadamard product.

Only the unmasked LLRs contribute to the training loss and are subsequently forwarded to the decoder during inference. This approach accommodates all modulation orders and enables the model to hierarchically learn bit significance, thereby enhancing the scalability of the receiver network across constellation shaping schemes with varying modulation orders.

\subsection{Loss Function and Model Training}

\begin{algorithm}[!t]
\caption{Training algorithm}\label{Alg1}
\begin{algorithmic}[1]
\Require Training data and channel samples, PAPR threshold $\epsilon_P$, initial Lagrangian multiplier $\lambda^{[0]}$ and penalty parameter $\mu^{[0]}$.
\Ensure The trained parameters $\mathbf{c}^*$, $\boldsymbol{\theta}^*$.
\State Initialize model parameters $\mathbf{c}$, $\boldsymbol{\theta}$.
  \For{$k=0,1,\dots,K-1$}
    \State \textbf{/* Perform multiple steps of SGD */} 
    \For{$t=0,1,\dots,T-1$}
        \State \textbf{Forward pass}: from $\mathbf{B}^{[t]}$ to $\mathbf{L}^{[t]}$
        \State \textbf{Compute}: $\mathcal{L}_{\mathrm{CE}}(\mathbf{c}, \boldsymbol{\theta})$, $\mathcal{L}_{\mathrm{P}}(\mathbf{c}, \epsilon_P)$
        \State \textbf{Compute gradients}: $\nabla_{\mathbf{c},\boldsymbol{\theta}}\mathcal{L}_{\mathrm{aug}}(\mathbf{c}, \boldsymbol{\theta}, \lambda^{[k]}, \mu^{[k]})$
        \State \textbf{Update parameters}: $\mathbf{c}$, $\boldsymbol{\theta}$
    \EndFor
    \State  \textbf{/* Update Lagrange multiplier */}
    \State \textbf{Recompute}: $\mathcal{L}_{\mathrm{P}}(\mathbf{\mathbf{c}}, \epsilon_P)$
    \State $\lambda^{[k+1]} \gets \lambda^{[k]} + \mu^{[k]}  \mathcal{L}_{\mathrm{P}}(\mathbf{\mathbf{c}}, \epsilon_P)$
    \State  \textbf{/* Update penalty parameter */}
    \State $\mu^{[k+1]}  \gets \tau \mu^{[k]}$, where $\tau > 1$\label{lst:eta}
  \EndFor
\end{algorithmic}
\end{algorithm}

\begin{figure*}[!t]
\centering
\includegraphics[width=0.8\textwidth]{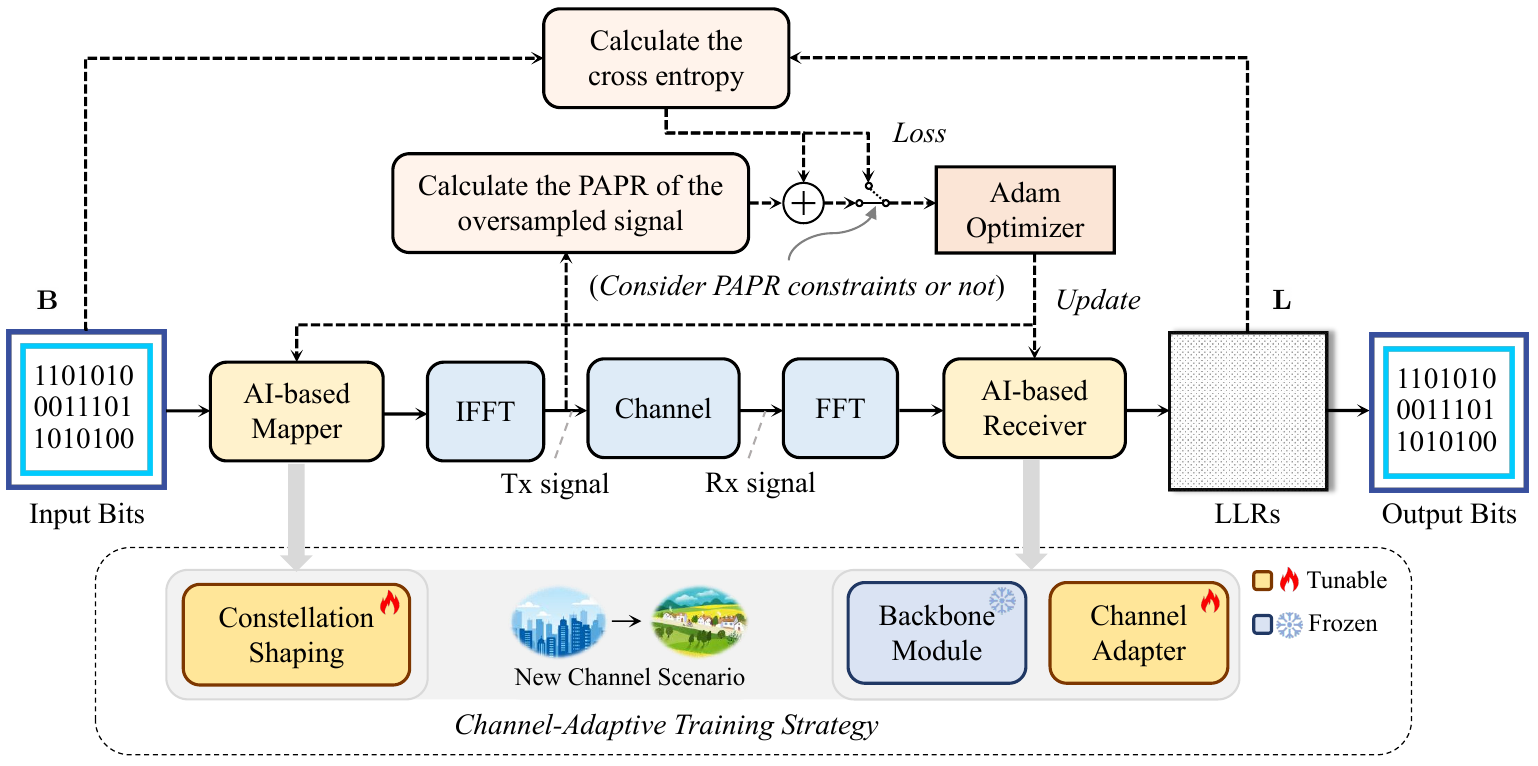}
\caption{Block diagram of the training process for the proposed end-to-end transceiver.}
\label{Fig: training diagram}
\end{figure*}

The overall training procedure of the proposed adaptive AI transceiver is illustrated in Fig.~\ref{Fig: training diagram}. The input to the transmitter is a randomly generated binary bitstream \(\mathbf{B}\), and the output of the receiver is the soft information \(\mathbf{L}\). The E2E system is trained through joint optimization of the constellation points $\mathbf{c}$ and the neural receiver parameters $\boldsymbol{\theta}$, which consist of the backbone module parameters $\boldsymbol{\theta}_{\text{backbone}}$ and the CA module parameters $\boldsymbol{\theta}_{\text{CA}}$. 

We adopt a composite loss function to train the E2E AI transceiver, which incorporates two key components: (i) the binary cross-entropy (CE) loss and (ii) the PAPR penalty. The CE loss measures the bit-level reconstruction accuracy and is defined as  
\begin{equation}  
\mathcal{L}_{\mathrm{CE}} = -\frac{1}{N_{b}}\sum_{i=1}^{N_{b}}\left(b_i \log(l_i) + (1 - b_i) \log(1 - l_i)\right),  
\end{equation}  
where \(b_i\) and \(l_i\) denote the ground-truth transmitted bit and the corresponding predicted LLR for the \(i\)-th bit, respectively. The total number of bits per training batch is given by \(N_b = MN_sN_c\).

To simultaneously suppress excessive PAPR and maintain a low BER, the optimization problem is formulated as
\begin{subequations}\label{optimization_problem_reform}
\begin{align}
&\underset{\mathbf{c}, \boldsymbol{\theta}}{\text{minimize}} \quad \mathcal{ L}_{\mathrm{CE}}({\mathbf{c}}, \boldsymbol{\theta}) \\
&\text{subject to} \quad \text{PAPR}(\mathbf{c}) \leq \epsilon_P, \label{papr_constraint_reform}
\end{align}
\end{subequations}
where $\epsilon_P$ denotes the target PAPR. However, directly counting the number of signal samples whose power exceeds the target peak value is a non-differentiable operation. To obtain a differentiable surrogate, the constraint in \eqref{papr_constraint_reform} can be equivalently expressed as
\begin{equation}
    \mathbb{E}\bigg( \text{max}\Big(\frac {{|\tilde{x}_n|}^2}{ \mathbb{E}[{|\tilde{x}_n|}^2]}-\epsilon_P, 0 \Big)\bigg)  = 0,
\end{equation}
The expectation can be approximated using Monte Carlo sampling of the transmit symbol, which is calculated as
\begin{align}\label{PAPR_const_MC}
    \mathcal{L}_{\mathrm{P}} = \frac{1}{B_sLN_c}\sum_{i=1}^{B_s}\sum_{n=1}^{LN_c} \text{max}\Big(\frac {{|\tilde{x}_n^{[i]} |}^2}{ \mathbb{E}[{|\tilde{x}_n^{[i]}|}^2]}-\epsilon_P, 0 \Big),
\end{align}
where $B_s$ denotes the batch size.

In this work, we employ the augmented Lagrangian method to solve the constrained optimization problem arising in the E2E transceiver design, inspired by~\cite{ait2022waveform}.
By constructing the augmented Lagrangian function, the original constraint formulation is transformed into an unconstrained problem, which allows the CE loss and the PAPR constraint to be jointly reformulated as a differentiable loss function. The augmented Lagrangian can be expressed as follows
\begin{align}\label{augmented_lagrangian}
\mathcal{L}_{\mathrm{aug}}(\mathbf{c}, \boldsymbol{\theta}, \lambda^{[k]}, \mu^{[k]})
&= \mathcal{L}_{\mathrm{CE}}(\mathbf{c}, \boldsymbol{\theta})
+ \lambda^{[k]} \mathcal{L}_{\mathrm{P}}(\mathbf{c}, \epsilon_P) \nonumber\\
&\quad + \frac{\mu^{[k]}}{2} \left| \mathcal{L}_{\mathrm{P}}(\mathbf{c}, \epsilon_P) \right|^2
\end{align}
where the superscript $[k]$ refers to the $k$-th iteration. $\lambda$ represents the Lagrangian multiplier for the PAPR constraint, and $\mu>0$ denotes the penalty parameter that is progressively increased. 
These factors serve as hyperparameters that balance the contributions of each loss component to the joint loss function. This mechanism prevents overemphasis on PAPR reduction that could otherwise distort constellation points and degrade detection accuracy.
The optimization is performed through stochastic gradient descent
(SGD) using the Adam~\cite{kingma2014adam} optimizer to compute gradients, followed by backpropagation through the system with respect to the trainable parameters. The strategy described in Algorithm~\ref{Alg1} has also been successfully applied to similar problems in~\cite{ait2022waveform}.

To enhance the adaptability of the AI-based transceiver in dynamic channel conditions while minimizing computational overhead, we adopt an online lightweight adaptation strategy as shown in Fig.~\ref{Fig: training diagram}. 
Specifically, this strategy decouples training into two phases: offline pre-training for generalizable feature extraction and online adaptation for real-time transmission. Here, the constellation points are fine-tuned to match the characteristics of the new channel environment. Freezing the backbone receiver network while updating only the constellation points and the parameters of the CA modules enables efficient transfer learning under limited data and resource constraints. The noise-aware mechanism in the CA module further ensures robustness against time-varying noise. This channel-adaptive training strategy ensures that the system remains both responsive and resource-efficient during real-time operation.

\section{Evaluations}

\subsection{Training and Evaluation Setup}
\begin{table}[!t]
\centering
\caption{System Parameters}
\label{tab: params}
\setlength{\tabcolsep}{3pt} 
\begin{tabular}{@{}>{\centering\arraybackslash}p{4.3cm}!{\vrule width 0.3pt}>{\centering\arraybackslash}p{2.5cm}@{}}
\toprule

\textbf{Parameter} & \textbf{Value} \\
\midrule

OFDM Symbols $N_s$ & 14 (1\,slot)\\
Subcarriers $N_c$ & 72 (6\,PRBs) \\
Receiving antennas $N_r$ & 32 \\
Carrier frequency & $3.5$\,GHz \\
Subcarrier spacing & $30$\,kHz \\
Slot duration & $0.5$\,ms \\
Delay spread & $100$\,ns \\
UE speed & $30, 120, 300$\,km/h \\
Channel coding scheme & LDPC \\
Batch size $B_s$ & 32 \\
Learning rate (training from scratch) & 0.001 \\
Learning rate (fine-tuning) & 0.0005 \\

\bottomrule
\end{tabular}
\end{table}

\begin{figure*}
	\centering
	
     \begin{subfigure}[b]{\linewidth}
         \centering
		\includegraphics[width=0.70\textwidth]{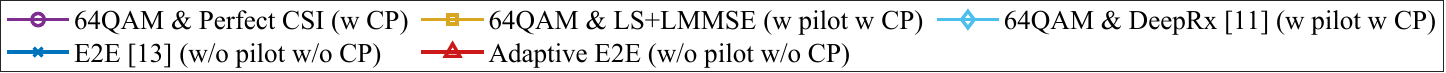}
     \end{subfigure}
	
    \vspace{0.15cm}
    
     \begin{subfigure}[b]{0.3\linewidth}
         \centering
		\includegraphics[width=1.0\textwidth]{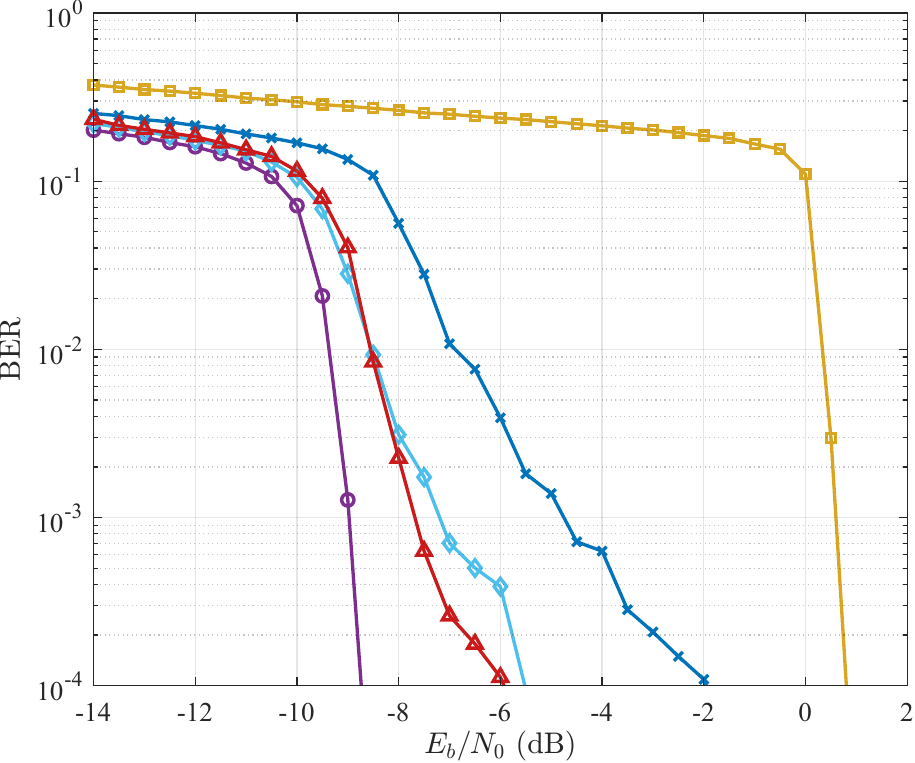}
         \caption{BER at $30$km/h.}\label{fig:ber_30}
     \end{subfigure}
     \begin{subfigure}[b]{0.3\linewidth}
         \centering
         \includegraphics[width=1.0\textwidth]{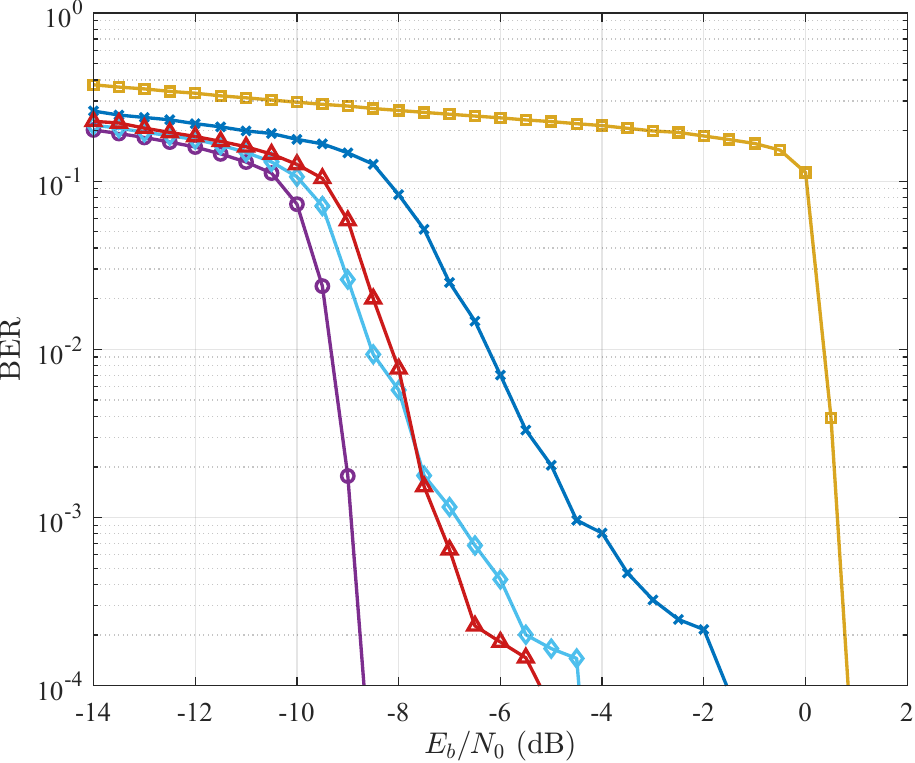}
         \caption{BER at $120$km/h.}\label{fig:ber_120}
     \end{subfigure}
     \begin{subfigure}[b]{0.3\linewidth}
         \centering
         \includegraphics[width=1.0\textwidth]{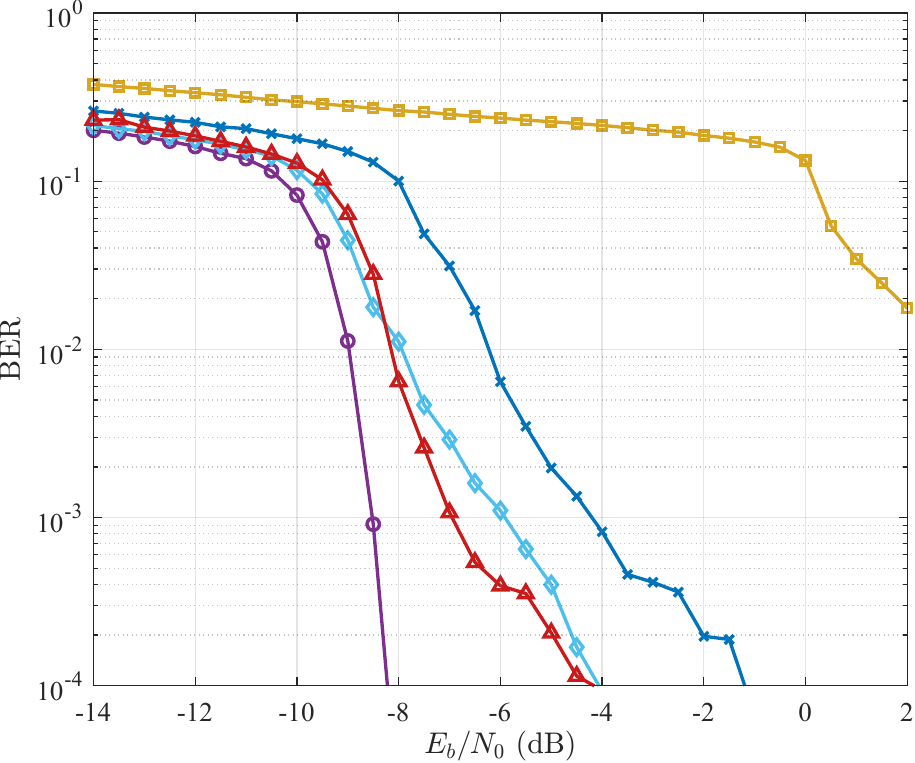}
         \caption{BER at $300$km/h.}\label{fig:ber_300}
     \end{subfigure}
	
     \begin{subfigure}[b]{0.3\linewidth}
         \centering
		\includegraphics[width=1.0\textwidth]{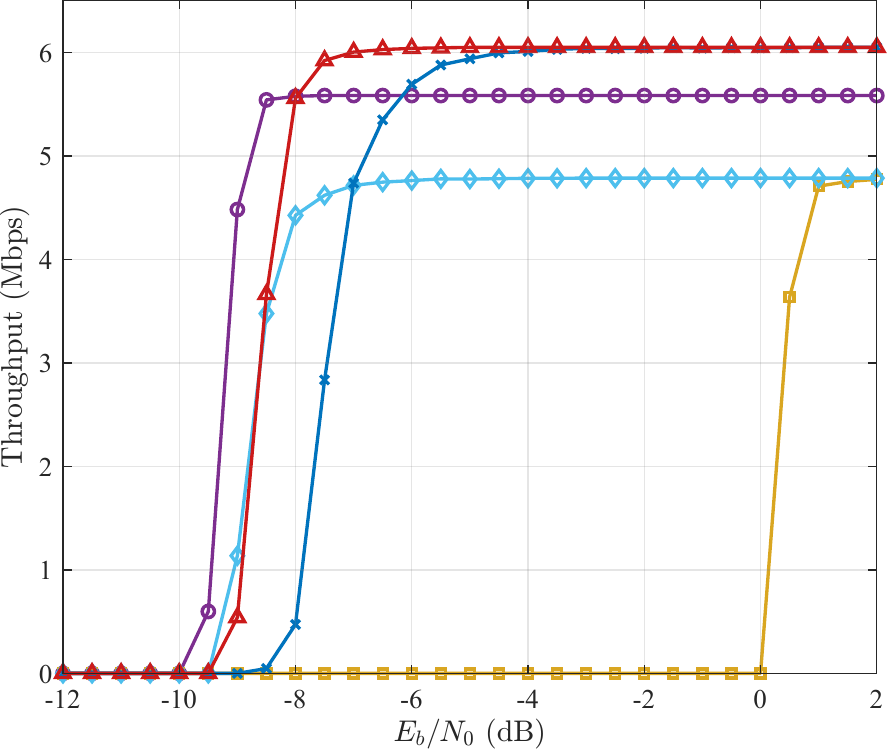}
         \caption{Throughput at $30$km/h.}\label{fig:throughput_30}
     \end{subfigure}
     \begin{subfigure}[b]{0.3\linewidth}
         \centering
         \includegraphics[width=1.0\textwidth]{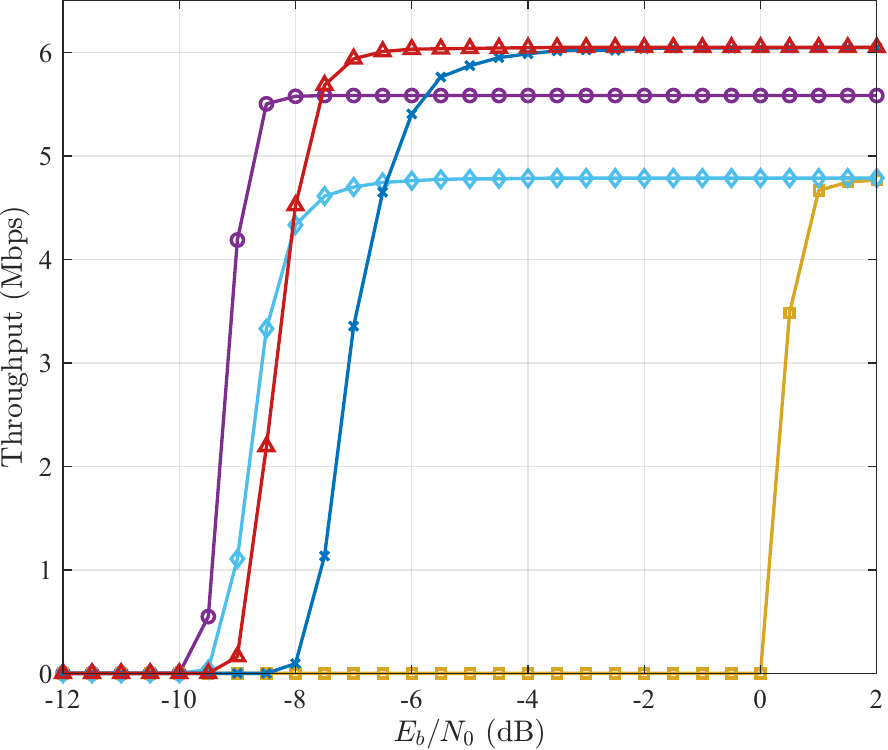}
         \caption{Throughput at $120$km/h.}\label{fig:throughput_120}
     \end{subfigure}
     \begin{subfigure}[b]{0.3\linewidth}
         \centering
         \includegraphics[width=1.0\textwidth]{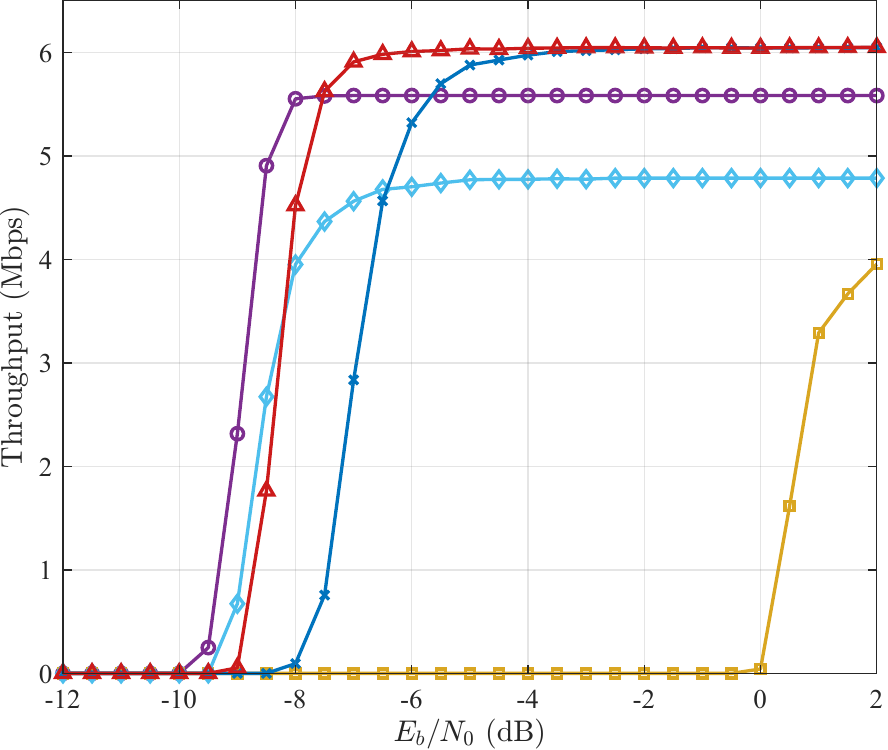}
         \caption{Throughput at $300$km/h.}\label{fig:throughput_300}
     \end{subfigure}
     \caption{BER and throughput performance of the evaluated schemes ($M=6$) in the CDL-C channel model for different speeds.\label{fig:cdlc_ber_throughput}}
\end{figure*}

\begin{figure}[!t]
\centering
\begin{subfigure}[b]{0.48\textwidth}
    \centering
        \includegraphics[width=0.84\textwidth]{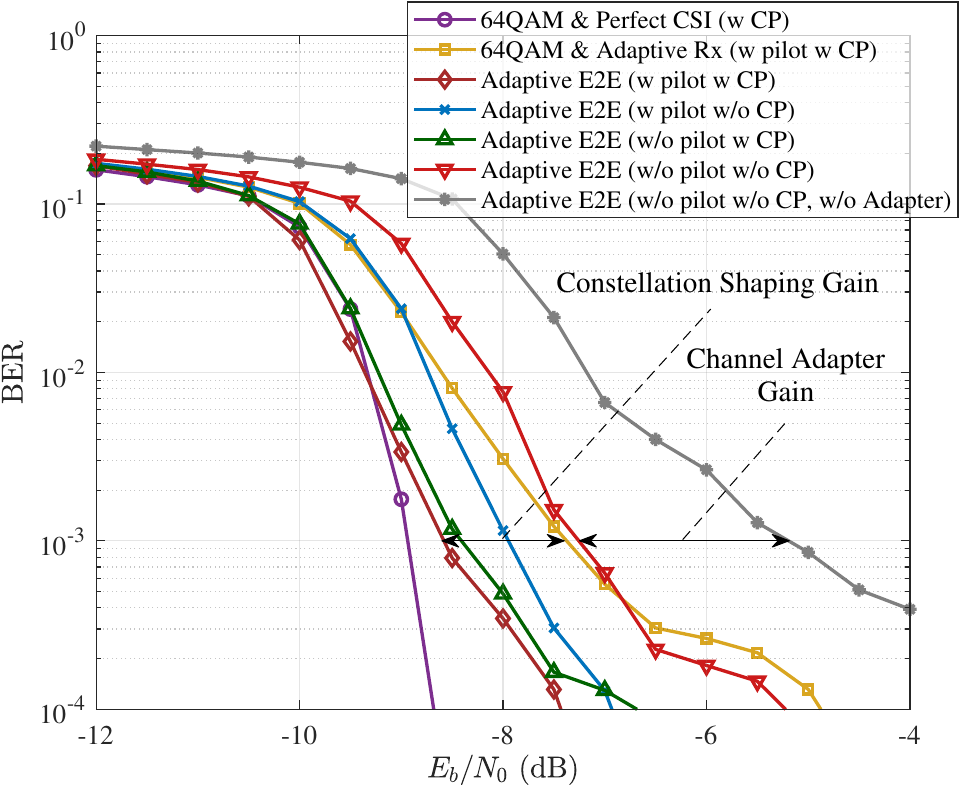}
    \caption{BER performance.}
    \label{Fig: ber_diff_config}
\end{subfigure}

\vspace{0.08cm} 

\begin{subfigure}[b]{0.48\textwidth}
    \centering
    \includegraphics[width=0.80\textwidth]{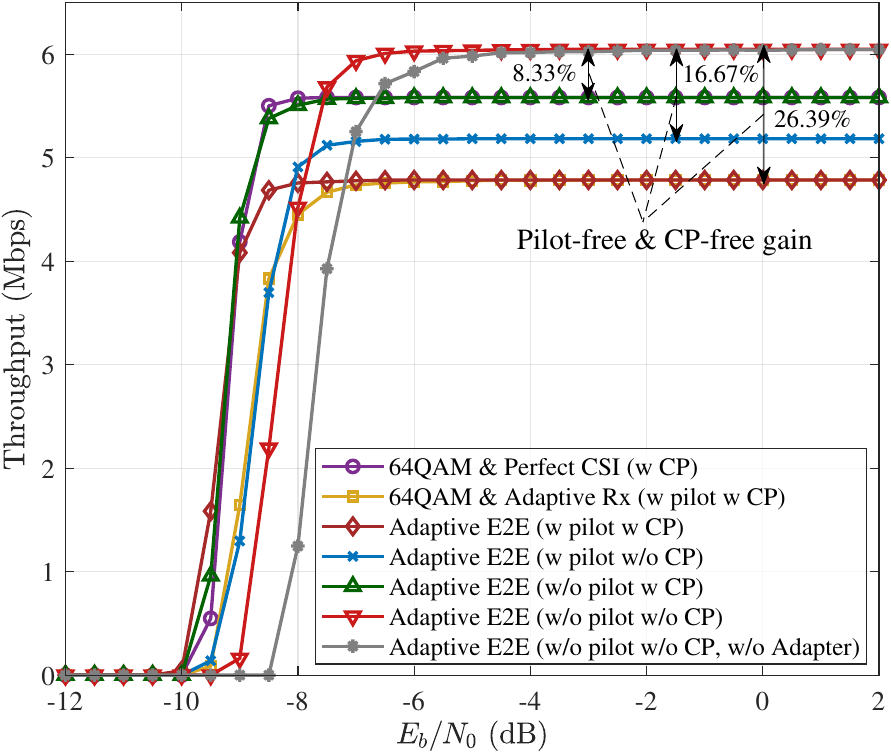}
    \caption{Throughput performance.}
    \label{Fig: throughput_diff_config}
\end{subfigure}

\caption{Comparison of proposed transceiver performance ($M=6$) under different pilot and CP configurations in the CDL-C channel model at $120$ km/h.}
\label{Fig: performance_diff_config}
\end{figure}

For realistic training and evaluation, the channel responses are generated using Sionna~\cite{hoydis2022sionna}. To demonstrate the effectiveness of our work, we present simulation results for the 3GPP cluster delay line (CDL) channel model and the 3GPP urban macro (UMa) channel model~\cite{3GPP38901}. The carrier frequency is set to $3.5$ GHz. Specifically, we consider a single-antenna UE transmitter and a BS receiver equipped with a $4 \times 4$ uniform planar antenna array with dual-polarized elements, resulting in $N_r=32$ receive antennas. The system operates in a single-stream configuration. A 5G NR-compliant low-density parity-check (LDPC) encoding and decoding are applied at a coding rate of $r=0.5$. For the system parameters, the OFDM system consists of $N_c=72$ subcarriers and $N_s=14$ OFDM symbols per slot. The conventional pilot-assisted baseline reserves two time-domain symbols per slot for pilot transmission and incorporates a 6-sample cyclic prefix to combat ISI and ICI (see Fig.~\ref{Fig: architecture}). Some of the simulation parameters used in this paper are listed in Table~\ref{tab: params}.

\begin{table}[!t]
\centering
\caption{Details of the Neural Network Architecture}
\label{tab:rx}
\renewcommand{\arraystretch}{1.2}
\setlength{\tabcolsep}{4pt}
\begin{tabular}{@{} L{2.1cm} !{\vrule width 0.5pt} C{1.6cm} !{\vrule width 0.5pt} C{1.6cm} !{\vrule width 0.5pt} C{1.8cm} @{}}
\toprule
\textbf{Layer Name} & \textbf{Filters/Units} & \textbf{Kernel Size} & \textbf{Dilation Rate} \\ 
\midrule
Input Conv2D        & 128  & (3,3)   & (1,1)   \\ 
Residual Block 1    & 128  & (7,7)   & (7,2)   \\ 
Residual Block 2    & 128  & (7,5)   & (7,1)   \\ 
Residual Block 3    & 128  & (5,3)   & (1,2)   \\ 
Residual Block 4    & 128  & (3,3)   & (1,1)   \\ 
Residual Block 5    & 128  & (3,3)   & (1,1)   \\
CA-DWConv           & 32  & (3,3)   & (1,1)   \\ 
CA-PWConv           & 128 & (1,1)   & (1,1)   \\ 
CA-AF-Dense 1   & 16   & --      & --      \\ 
CA-AF-Dense 2   & 128  & --      & --      \\ 
Output Conv2D       & $M$/$M_{\text{max}}$ & (1,1)   & (1,1)   \\ 
\bottomrule
\end{tabular}
\begin{tablenotes}
\footnotesize
\item[*] *$M$: single-modulation-order training\\
\item[] $M_{\text{max}}$: multi-modulation-order training
\end{tablenotes}
\end{table}

The parameters of the transmitter network are the trainable constellation points, so the number of parameters depends on the modulation order. At the receiver side, we set the number of Residual blocks and CA modules in the network to $N_L=5$. The channel reduction ratio $\gamma$ in the CA module can be adaptively adjusted based on the computational capacity of resource-constrained devices. In this work, $\gamma$ is set to $4$.
Detailed information for each layer of the receiver network can be found in Table~\ref{tab:rx}.

All AI-based methods are trained with a total of 30,000 parameter updates under identical hardware settings and hyperparameter configurations to ensure a fair comparison. Specifically, for the methods considering the PAPR constraint, the training is organized into $K=2500$ outer iterations, each comprising $T=12$ inner steps of SGD. The Lagrange multiplier is initialized as $\lambda^{[0]} = 0$ and the penalty parameter is initialized as $\mu^{[0]} = 0.1$, with the penalty scaling factor set to $\tau = 1.004$. In contrast, methods trained solely based on the CE loss adopt their original single-loop training procedures. To ensure computational fairness, these methods are also trained for a total of 30,000 iterations.

To demonstrate the effectiveness of the proposed approach, we evaluate and compare its performance against other methods in terms of BER and throughput. The throughput metric is defined as
\begin{equation} \text{Throughput} = N_{\text{slot}} N_{\text{RE}} r\rho M(1-\text{BLER}), \label {x}\end{equation}
where $N_{\text{slot}}$ is the number of slots per second, $N_{\text{RE}}=N_sN_c$ is the number of REs forming a slot, and $\rho$ is the ratio of REs carrying data symbols.
In our evaluation, the pilot-free and CP-free scheme achieves the highest data resource utilization with \(\rho = 1\). For comparison, the pilot-aided scheme without CP has \(\rho = 6/7\), the CP-aided scheme without pilots has \(\rho = 12/13\), and the conventional pilot-aided and CP-aided system has \(\rho = 72/91\).

\subsection{Performance Comparison of Different Methods}

We conduct a comprehensive comparison with several benchmark schemes. Specifically, the proposed method is evaluated against: (i) traditional QAM modulation and the ideal case assuming perfect CSI at the receiver; (ii) QAM modulation and a conventional receiver employing LS channel estimation and LMMSE equalization; (iii) QAM modulation and a neural receiver, namely DeepRx~\cite{honkala2021deeprx}, which leverages pilots and CP for prior channel estimation; and (iv) a fully E2E learning-based transceiver without pilots and CP~\cite{aoudia2021end,ait2021trimming}. We evaluate the BER and throughput performance of the proposed method in the CDL-C channel model across three different speeds and a range of \( E_b/N_0 \) values. All AI-based methods are jointly trained over mixed samples with \( E_b/N_0 \) uniformly distributed between \(-10\) dB and \(5\) dB to ensure robustness across varying channel conditions. 

As shown in Fig.~\ref{fig:cdlc_ber_throughput}, the perfect CSI scenario consistently achieves the lowest BER across the entire range of $E_b/N_0$, owing to the assumption that the receiver has prior knowledge of the channel matrix, thereby eliminating channel estimation errors. However, it is worth noting that even under the ideal perfect CSI assumption, a CP is still required to mitigate ISI and ICI. As a result, the proposed method achieves higher throughput than the perfect CSI case when \( E_b/N_0 > -8\ \)dB, owing to its CP-free design, which avoids the overhead introduced by the cyclic prefix and demonstrates improved spectral efficiency under practical channel conditions. Moreover, it can be observed that the baseline schemes based on LS channel estimation and LMMSE equalization consistently result in the highest BER across all speeds. In addition, their BER performance deteriorates significantly as the speed increases. The DeepRx receiver achieves slightly better BER performance than the proposed method at low $E_b/N_0$. This is attributed to its prior explicit channel estimation using pilot signals, as well as its CP-based transceiver architecture, both of which provide improved robustness in high-noise conditions. However, the inclusion of pilot signals and the cyclic prefix also leads to a significant reduction in spectral efficiency. Notably, at higher \( E_b/N_0 \) regimes, the proposed pilot-free and CP-free transceiver achieves even lower BER than the pilot-aided and CP-aided systems. The observed performance gains demonstrate the effectiveness of combining a neural receiver with a trainable custom constellation and provide valuable insights for future practical deployments and standardization efforts.

Furthermore, as illustrated in Fig.~\ref{fig:cdlc_ber_throughput}, it can be observed that the proposed adaptive E2E network consistently provides at least a \(2.5\) dB performance gain across different speeds at the BER level of \(10^{-3}\), compared to the existing E2E design in~\cite{aoudia2021end}, which similarly operates without pilot signals and CP. 
In addition, under low \(E_b/N_0\) conditions of \(-12\)~dB with a user mobility of 120~km/h, the proposed method achieves a BER of \(1.84\times10^{-1}\) while the existing E2E network yields \(2.20\times10^{-1}\), representing an approximate $16.36\%$ performance gain.
This performance improvement is primarily attributed to the inserted channel adapter module with a bottleneck structure. This design enhances the extraction of implicit spatio-temporal-frequency channel features from the data resource blocks. Besides, it leverages noise power as auxiliary information, thereby improving the network's robustness to noise.

\begin{figure}[!t]
\centering
\includegraphics[width=0.40\textwidth]{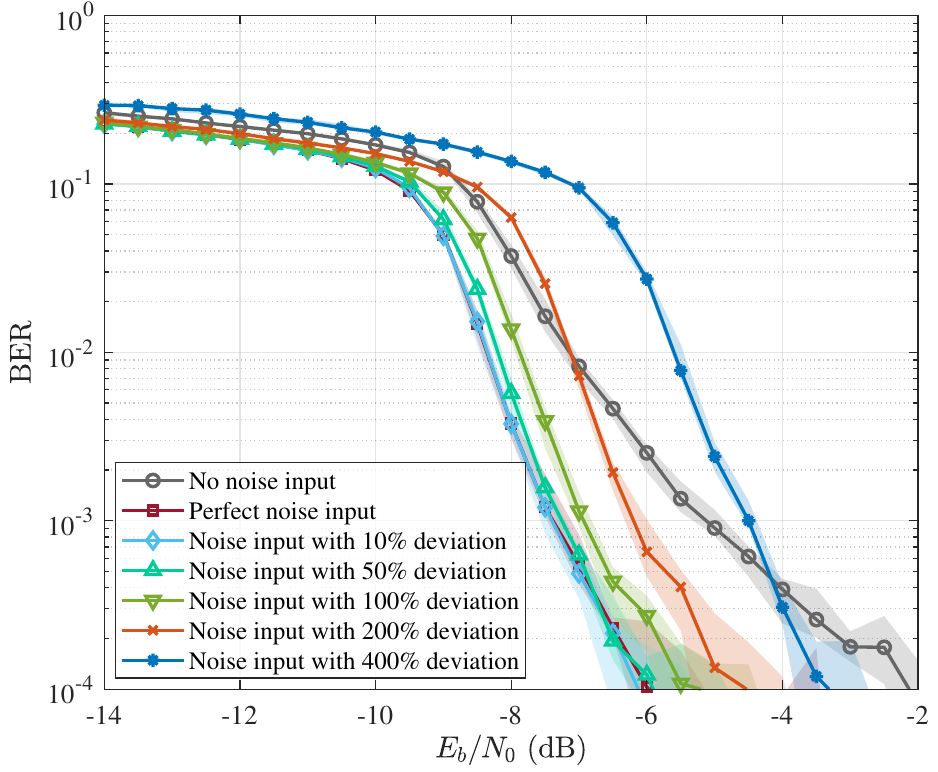}
\caption{BER comparison of the proposed pilot-free and CP-free transceiver under noise mismatch in the CDL-C channel model at $120$ km/h. The dark solid line represents the median BER over six random trials, while the shaded region indicates the range between the minimum and maximum BER. }
\label{Fig: noise_ablation}
\end{figure}

In addition, Fig.~\ref{Fig: performance_diff_config} presents a performance comparison of the proposed transceiver operating under different pilot and CP configurations. Specifically, Fig.~\ref{Fig: performance_diff_config}(\subref{Fig: ber_diff_config}) compares the BER performance. It can be observed that the AI-based constellation shaping provides approximately a \(1.16\) dB gain at a BER of $10^{-3}$ compared to conventional QAM modulation, demonstrating the advantage of non-uniform geometric shaping. In addition, the inclusion of the channel adapter yields an additional gain of about \(2\)~dB at a BER of \(10^{-3}\) compared with the configuration without the adapter. The CA module is lightweight and introduces negligible computational overhead. Specifically, the receiver with the adapter exhibits only a slight increase in computational cost, with the number of floating point operations (FLOPs) rising from $7.812$~GFLOPs to $8.227$~GFLOPs. Meanwhile, although the pilot-free and CP-free configuration shows a slight BER degradation compared to those relying on pilots or CP, it achieves a notable throughput improvement, as shown in Fig.~\ref{Fig: performance_diff_config}(\subref{Fig: throughput_diff_config}). In particular, it delivers a $26.39\%$ gain over the configuration that incorporates both pilot and CP. Notably, the learned constellations remain decodable by conventional receivers when pilots are available, 
whereas under the pilot-free configuration, reliable symbol recovery depends on the neural receiver.

In previous experiments, the proposed network is assumed to have access to the perfect noise power. However, in practical systems, the noise power is typically estimated from the uplink sounding reference signal (SRS) or other auxiliary mechanisms, and obtaining an accurate noise estimate is challenging.
In Fig.~\ref{Fig: noise_ablation}, we evaluate the BER performance of the proposed model when the input noise power deviates from the true value by different levels, in order to demonstrate the robustness of the method. 
It can be observed that when the input noise deviates by $10\%$ or $50\%$ from the true noise power, the model performance remains nearly unchanged. Even under $100\%$ deviation, the degradation is only about $0.5$ dB. A noticeable drop of approximately $2.5$ dB occurs only when the deviation reaches $400\%$. Furthermore, an ablation comparison with and without noise input shows a $2.2$ dB gain at a BER of $10^{-3}$, confirming the importance of incorporating noise power as an auxiliary prior in the proposed framework.

To analyze the computational complexity, we have measured the average running time of both the proposed transceiver and the conventional transceiver on a Windows server equipped with an Intel i7-7700K CPU and an NVIDIA GTX 1080Ti GPU. The average running time of the proposed transceiver is approximately $7.02\times10^{-2}$~seconds for each resource grid, compared with $8.21\times10^{-2}$~seconds for the conventional transceiver employing LS-based channel estimation and LMMSE equalization.

\subsection{Adaptation to New Environments}
\begin{figure}[!t]
\centering
\begin{subfigure}[b]{0.48\textwidth}
    \centering
    \includegraphics[width=0.82\textwidth]{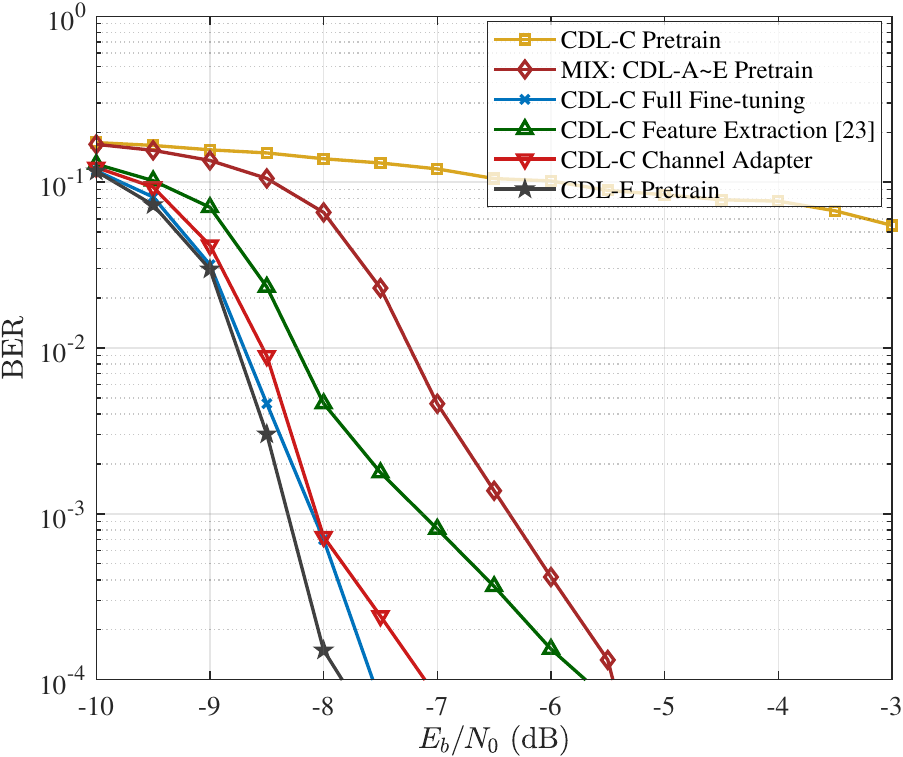}
    \caption{Evaluation in the CDL-E channel model.}
    \label{Fig: CDL-E BER}
\end{subfigure}

\vspace{0.08cm}  

\begin{subfigure}[b]{0.48\textwidth}
    \centering
    \includegraphics[width=0.82\textwidth]{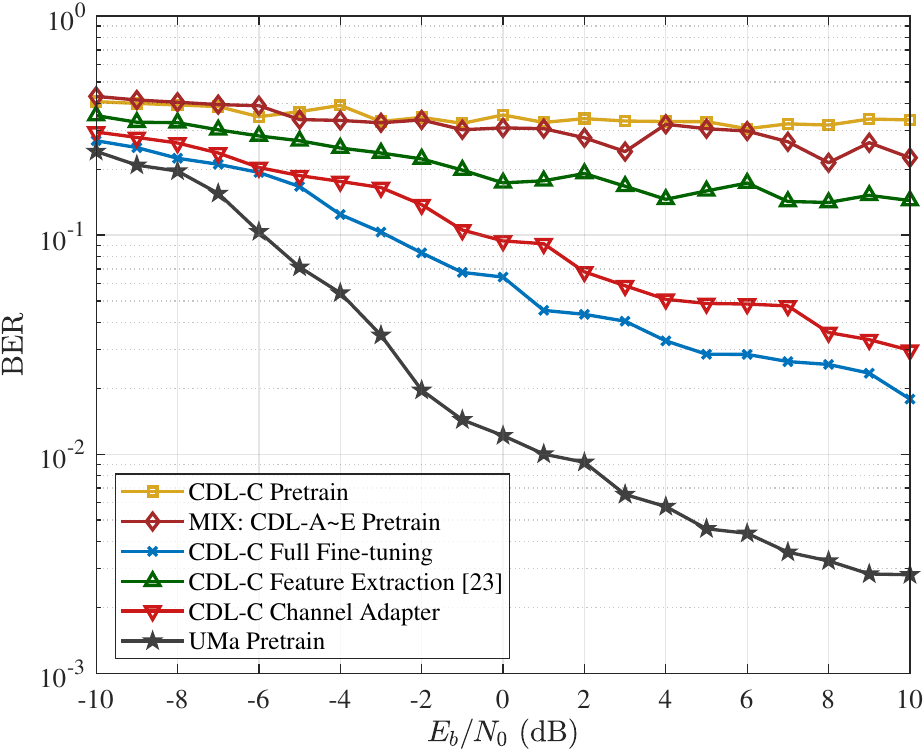}
    \caption{Evaluation in the UMa channel model.}
    \label{Fig: UMa BER}
\end{subfigure}

\caption{BER comparison between the proposed pilot-free and CP-free transceiver with channel adapter and baseline methods in the CDL-E and UMa channel models at $120$ km/h.}
\label{Fig: adapter_cdle_uma}
\end{figure}

The performance variation of the proposed adaptive E2E transceiver with channel adapter under mismatched channel conditions is also evaluated using two transfer learning strategies: full fine-tuning and the feature extraction method proposed in~\cite{uyoata2024transfer}. 
In the latter approach, all weights from the source model are transferred to the target model and frozen, while only the newly added one more ResNet layer and output Conv2D layer are fine-tuned on the target dataset. The additional layer also introduces extra computational and storage overhead during inference.
For all strategies, the channel sample dataset and the number of training epochs used for fine-tuning are set to $25\%$ of those used in the pretraining stage. In addition, we include several baseline networks pretrained under different conditions: CDL-C, a mixture of CDL-(A–E) channel models, and the target channel model. 

Fig.~\ref{Fig: adapter_cdle_uma} compares the BER of different approaches across various $E_b/N_0$ values in the CDL-E and UMa channel models. It can be observed that the model pretrained on the CDL-C channel without fine-tuning exhibits the worst BER performance, followed by the one trained on mixed CDL-(A–E) scenarios. The generalization capability of the model trained on mixed CDL scenarios is limited, particularly when applied to the UMa channel, where a noticeable performance gap is observed. In contrast, our proposed adapter-based fine-tuning method achieves performance comparable to that of full fine-tuning, while significantly outperforming the feature extraction method, which also updates only a subset of the model parameters.

\begin{table}[!t]
\centering
\caption{Comparison of Different Transfer Learning Methods}
\label{tab:adapter_comparison}
\renewcommand{\arraystretch}{1.3}
\setlength{\tabcolsep}{8pt}
\begin{tabular}{@{} >{\raggedright\arraybackslash}m{3.0cm} 
                !{\vrule width 0.5pt} 
                >{\centering\arraybackslash}m{1.3cm} 
                !{\vrule width 0.5pt} 
                >{\centering\arraybackslash}m{1.1cm} 
                >{\centering\arraybackslash}m{1.1cm} @{}}
\toprule
\multicolumn{1}{@{}>{\raggedright\arraybackslash}m{2.95cm}!{\vrule width 0.5pt}}{\multirow{2}{*}{\textbf{Methods}}} & 
\multicolumn{1}{c!{\vrule width 0.5pt}}{\multirow{2}{*}{\makecell{\textbf{Trainable} \\ \textbf{Params}}}} & 
\multicolumn{2}{c@{}}{\textbf{Average BER}} \\
\cmidrule(l{0.3em}r{0.3em}){3-4}
 &  & \textbf{CDL-E} & \textbf{UMa} \\
\midrule
Full Fine-Tuning & 6.49M & 0.01178 & 0.09860 \\ 
Feature Extraction~\cite{uyoata2024transfer} & 0.81M & 0.01660 & 0.21588 \\
\rowcolor{gray!10}
\textbf{Channel Adapter (Ours)} & 0.23M & 0.01338 & 0.12650 \\
\bottomrule
\end{tabular}
\end{table}

Table~\ref{tab:adapter_comparison} summarizes the performance and model complexity of different transfer learning strategies evaluated in the CDL-E and UMa channel models. The full fine-tuning approach yields the best BER performance across both channel models, albeit at the cost of retraining the entire network, which leads to substantial computational and storage overhead.
In contrast, the feature extraction method~\cite{uyoata2024transfer}, which fine-tunes an additional ResNet block on top of a frozen backbone, reduces training complexity but still incurs higher inference latency and memory usage compared to our proposed approach. Moreover, it suffers from significant performance degradation, especially in the UMa channel.
By comparison, the proposed channel adapter method achieves a favorable trade-off between efficiency and performance by introducing only around $3.5\%$ of full fine-tuning parameters for the adaptive E2E model. Notably, the proposed channel adapter exhibits substantial performance gains over the feature extraction method, achieving a relative BER reduction of $19.40\%$ and $41.40\%$ in the CDL-E and UMa channel models, respectively. These results highlight the effectiveness of the proposed lightweight adaptation mechanism in facilitating robust and efficient transfer across diverse channel environments. 

\subsection{Scalability to Multiple Modulation Orders}
\begin{figure*}[!t]
\centering
     \begin{subfigure}[b]{0.245\linewidth}
         \centering
		\includegraphics[width=0.97\textwidth]{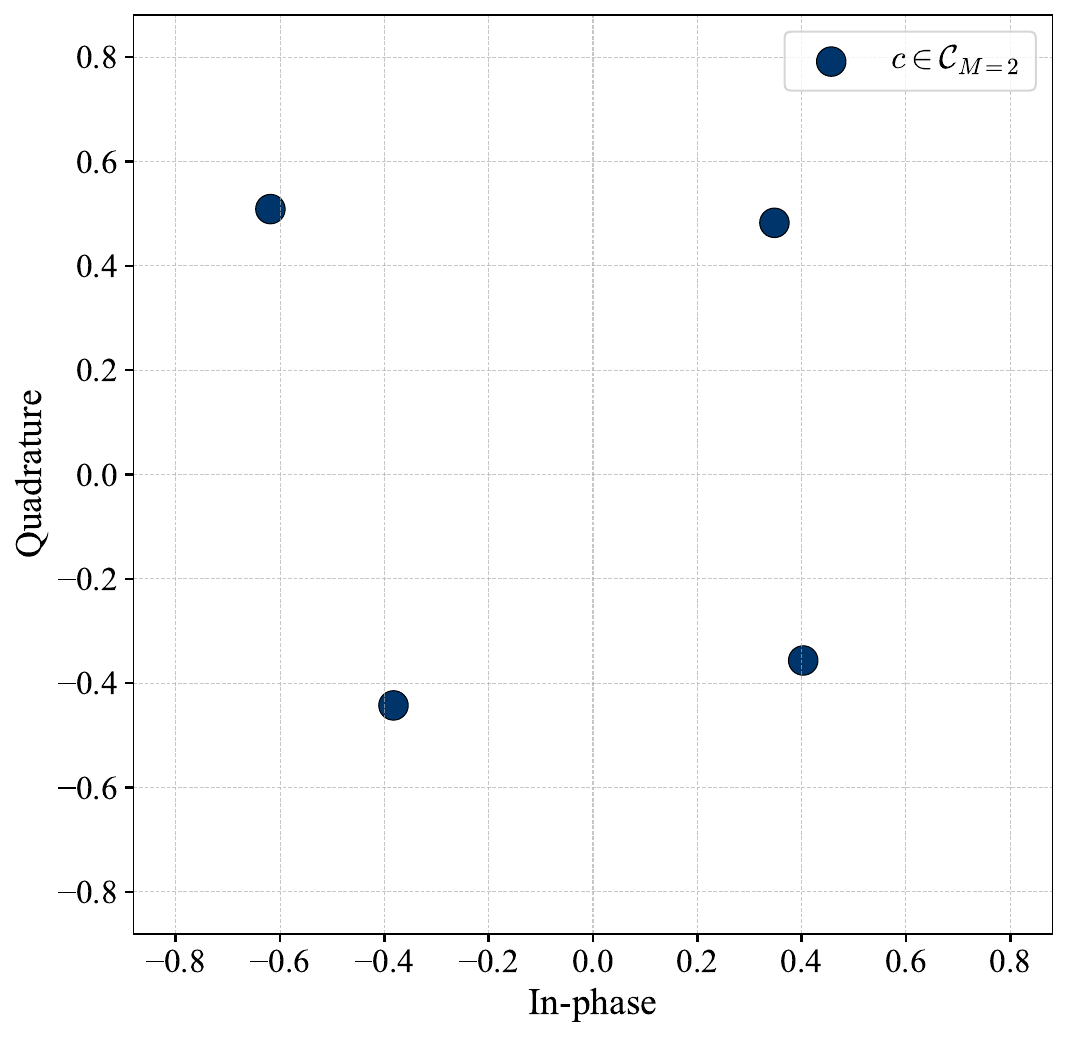}
         \caption{$M=2$.}
         \label{Fig: constellation_m2}
     \end{subfigure}
     \begin{subfigure}[b]{0.245\linewidth}
         \centering
		\includegraphics[width=0.97\textwidth]{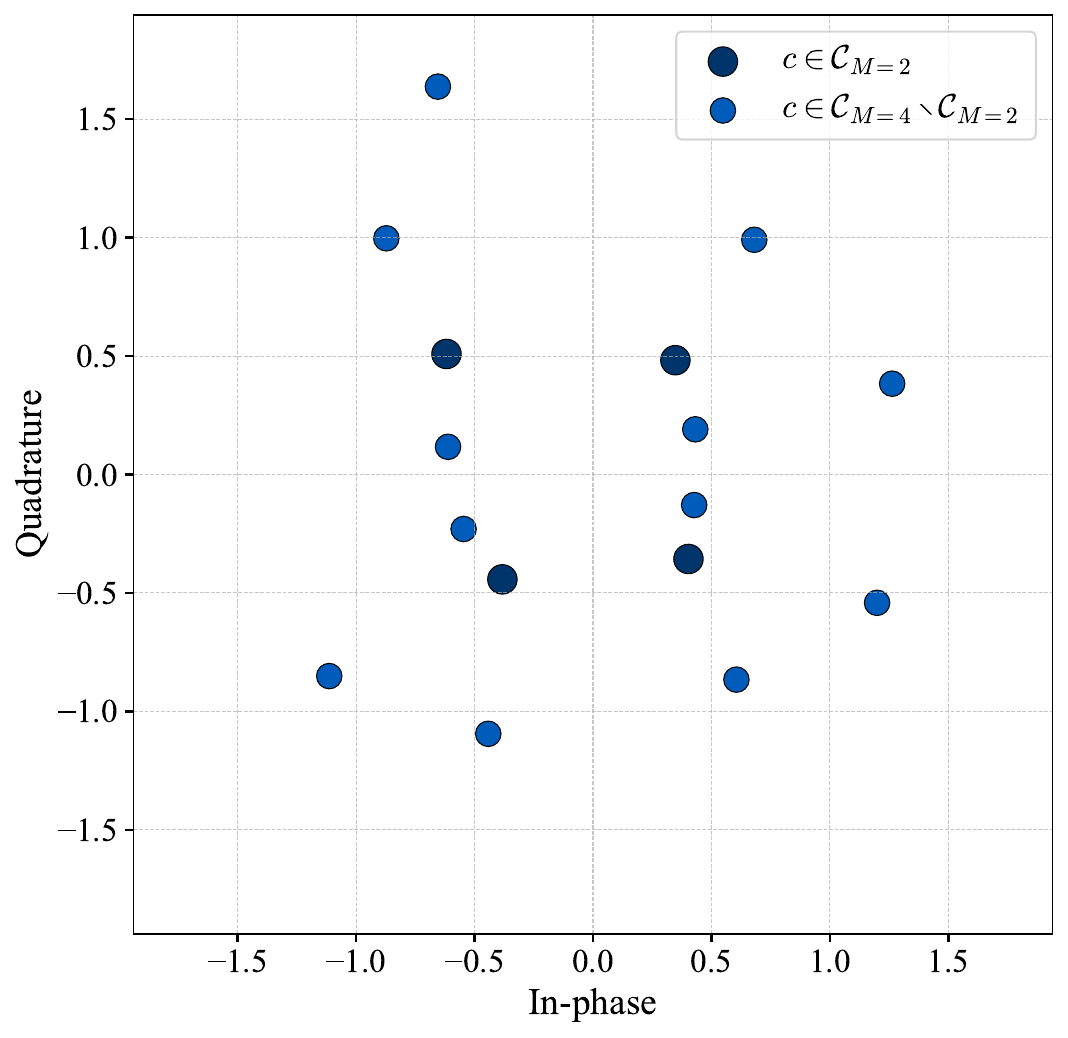}
         \caption{$M=4$.}
         \label{Fig: constellation_m4}
     \end{subfigure}
     \begin{subfigure}[b]{0.245\linewidth}
         \centering
		\includegraphics[width=0.95\textwidth]{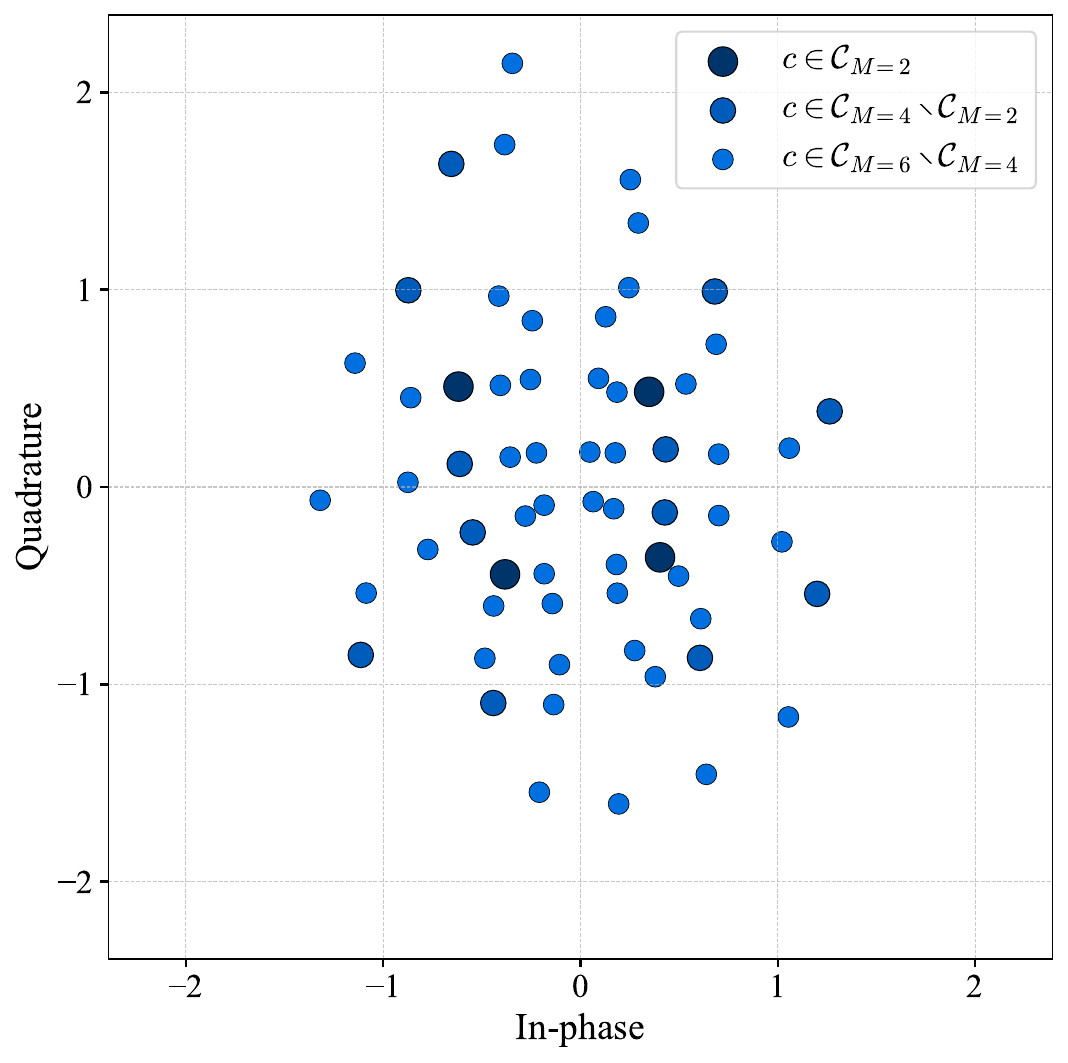}
         \caption{$M=6$.}
         \label{Fig: constellation_m6}
     \end{subfigure}
     \begin{subfigure}[b]{0.245\linewidth}
         \centering
		\includegraphics[width=0.95\textwidth]{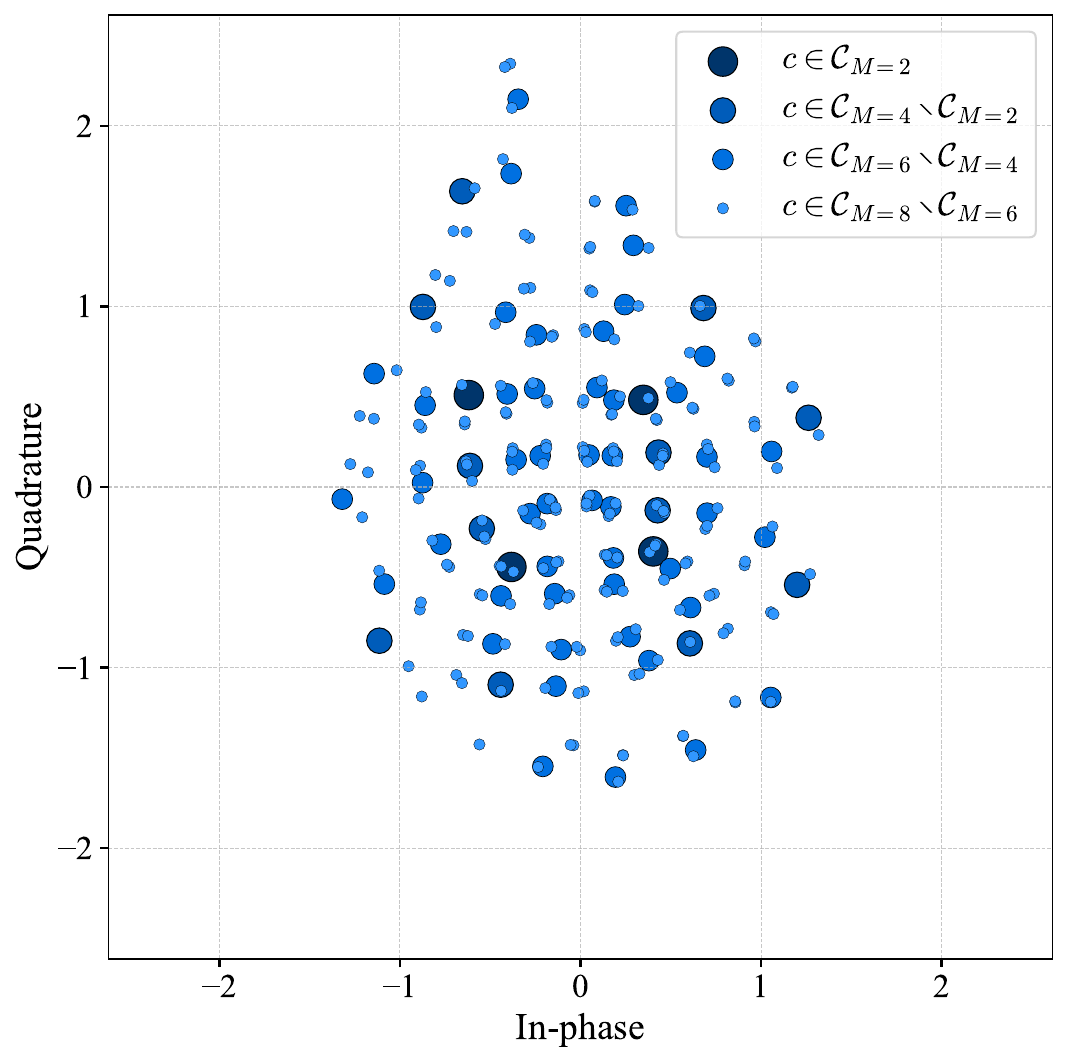}
         \caption{$M=8$.}
         \label{Fig: constellation_m8}
     \end{subfigure}

\caption{Learned constellation points from a unified architecture across multiple modulation orders.}
\label{Fig: constellation_multiorder}
\end{figure*}

\begin{figure}[!t]
\centering
\begin{subfigure}[b]{0.48\textwidth}
    \centering
    \includegraphics[width=0.80\textwidth]{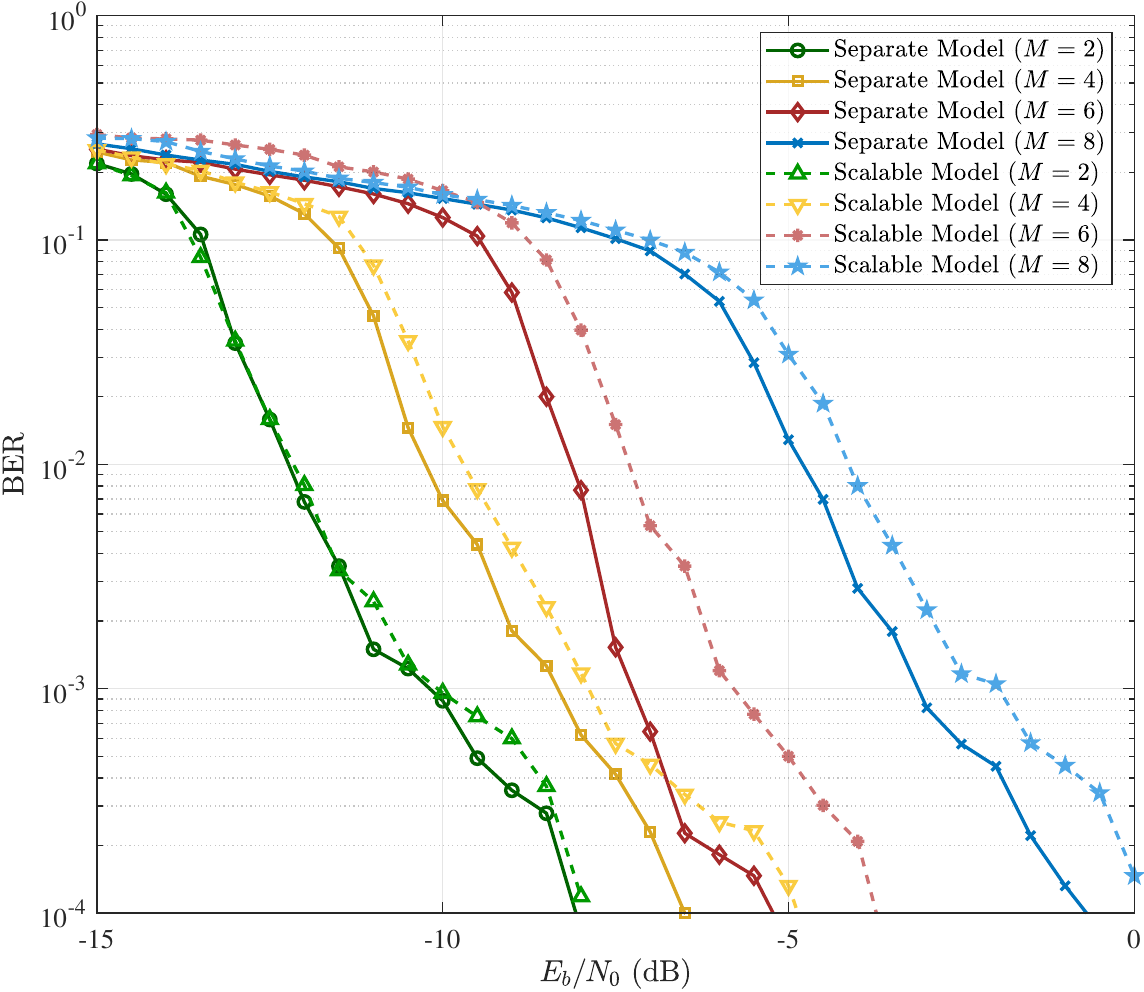}
    \caption{BER comparison of the proposed scalable transceiver.}
    \label{Fig: ber_Multi-order}
\end{subfigure}

\vspace{0.08cm} 

\begin{subfigure}[b]{0.48\textwidth}
    \centering
    \includegraphics[width=0.80\textwidth]{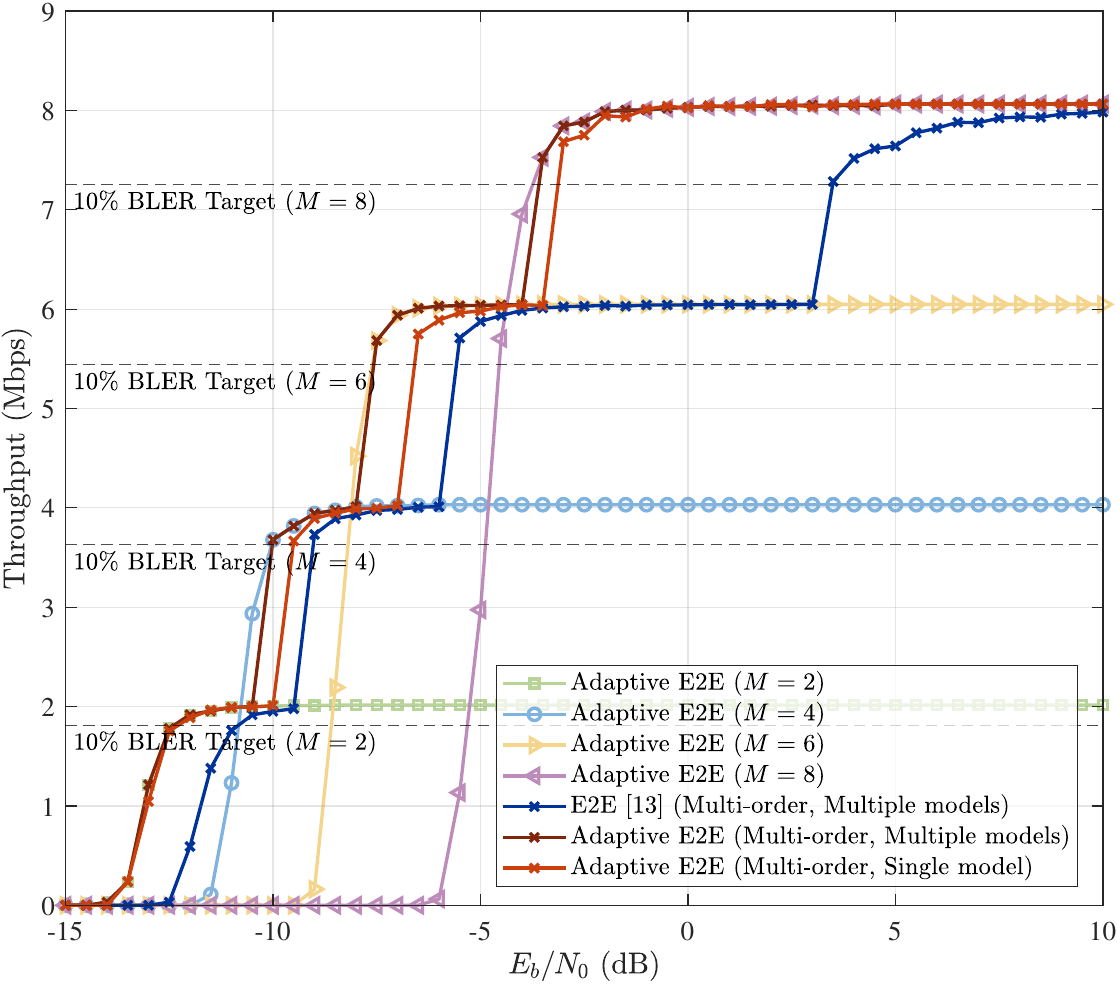}
    \caption{Throughput comparison of different schemes.}
    \label{Fig: throughput_Multi-order}
\end{subfigure}

\caption{BER and throughput comparison under pilot-free and CP-free configuration across modulation orders in the CDL-C channel model at $120$ km/h.}
\label{Fig: Multi-order}
\end{figure}

Fig.~\ref{Fig: constellation_multiorder} presents the learned constellation points obtained from the unified architecture across different modulation orders. It can be observed that each lower-order modulation constellation forms a subset of the higher-order one. Based on this hierarchical structure, the receiver performs demodulation by applying a masking operation to the output LLRs.
Furthermore, by observing the learned constellations, one or several constellation points deviate from most of the others, implicitly acting as “anchor” symbols for capturing channel characteristics. The remaining points exhibit non-uniform geometric shaping. This further illustrates how geometric shaping can be leveraged to empower pilot-free and CP-free systems. Additionally, the learned constellation points approximately follow Gray labeling, which is omitted from the figure for clarity.

The scalability performance of the proposed transceiver over different modulation orders in the CDL-C channel is presented in Fig.~\ref{Fig: Multi-order}. In this experiment, the modulation orders are set to $M=\{2,4,6,8\}$ for comparison. In Fig.~\ref{Fig: Multi-order}(\subref{Fig: ber_Multi-order}), the scalable model is trained on a mixed dataset with the maximum modulation order of $M_{\text{max}}=8$, while the separate model is trained individually for each specific modulation order without employing the proposed scalability mechanism. It can be observed that the scalable transceiver achieves comparable performance to the model trained for specific modulation orders. In addition, Table~\ref{tab: storage_overhead} compares the model storage overhead, showing that the scalable design reduces the overall storage requirement by $75\%$ compared with the separate models.

\begin{table}[!t]
\centering
\caption{Comparison of Model Storage Overhead}
\label{tab: storage_overhead}
\setlength{\tabcolsep}{4pt} 
\begin{tabular}{@{}>{\centering\arraybackslash}p{3.3cm}!{\vrule width 0.3pt}>{\centering\arraybackslash}p{1.4cm}!{\vrule width 0.3pt}>{\centering\arraybackslash}p{1.4cm}!{\vrule width 0.3pt}>{\centering\arraybackslash}p{1.7cm}@{}}
\toprule
\textbf{Model Type} & \textbf{Tx Params} & \textbf{Rx Params} & \textbf{Total Params} \\
\midrule
Separate - \{$M={2,4,6,8}$\}  & 680  & 102.4 M & 102.4 M\\
\rowcolor{gray!10}
\textbf{Scalable}    & 512  & 25.6 M  & 25.6 M \\
\bottomrule
\end{tabular}
\end{table}

\begin{figure}[!t]
\centering
\includegraphics[width=0.42\textwidth]{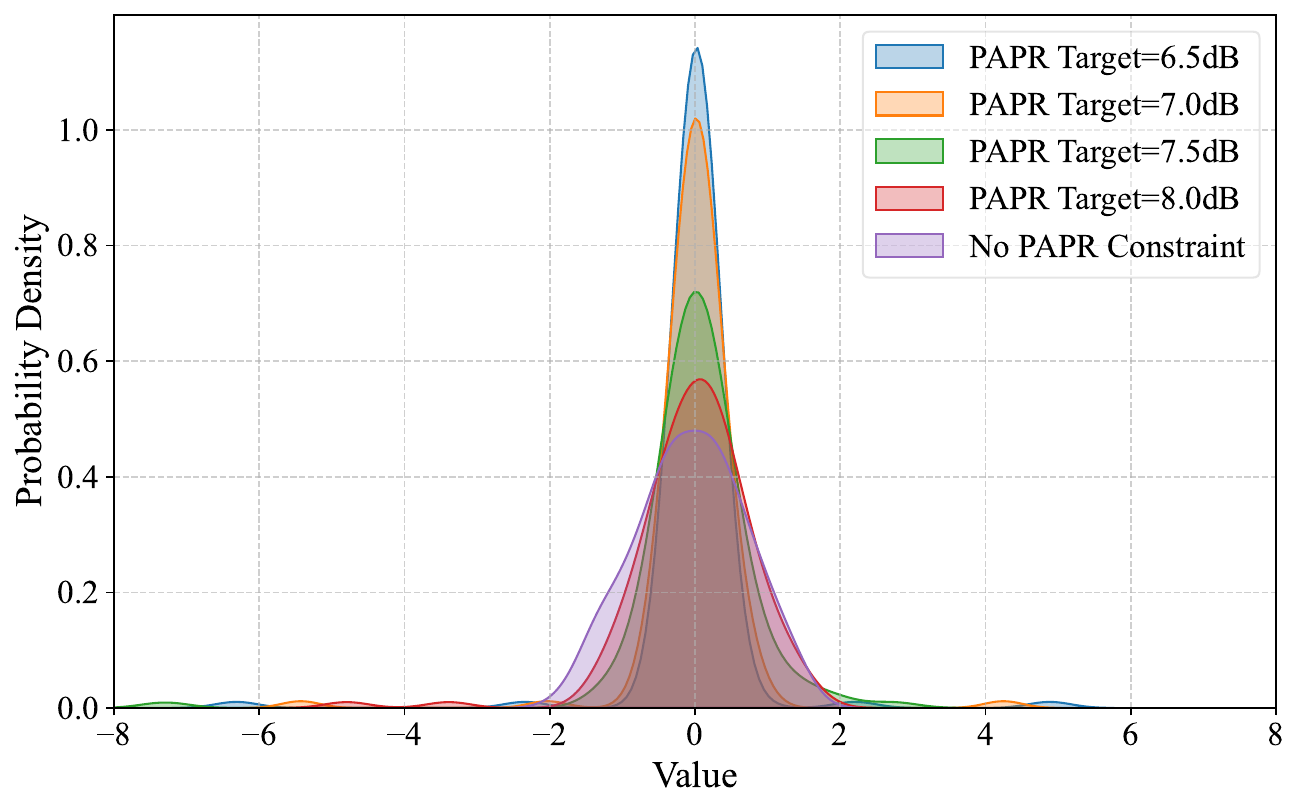}
\caption{Comparison of geometric distribution of constellation points (combined in-phase and quadrature components) under different PAPR constraints.}
\label{Fig: papr_kde}
\end{figure}

In the mixed training with multiple modulation orders, the modulation order at each $E_b/N_0$ is selected as the highest one that achieves a BLER target of $10\%$. Fig.~\ref{Fig: Multi-order}(\subref{Fig: throughput_Multi-order}) illustrates the throughput performance of the proposed pilot-free and CP-free transceiver across different modulation orders in the CDL-C channel model. The proposed modulation-order scalable strategy incurs only a slight performance degradation at higher modulation orders compared to deploying separate models for each order. Moreover, the proposed scalable transceiver for multiple modulation orders exhibits significant performance gains over the E2E approach in~\cite{aoudia2021end}, which employs separate models for each modulation level. Notably, when targeting a BLER of $10\%$ at \(M=8\), our method achieves an approximate $6.5$ dB improvement in performance. In addition, the unified model operates seamlessly across different modulation orders without model switching and avoids additional latency overhead, which is beneficial for real-time NextG applications.

\subsection{Performance of PAPR-Constrained Transmission}

\begin{figure*}[!t]
\centering
	
     \begin{subfigure}[b]{\linewidth}
         \centering
		\includegraphics[width=0.95\textwidth]{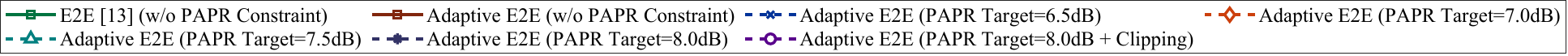}
     \end{subfigure}
    
    \vspace{0.08cm}
     \begin{subfigure}[b]{0.495\linewidth}
         \centering
		\includegraphics[width=0.9\textwidth]{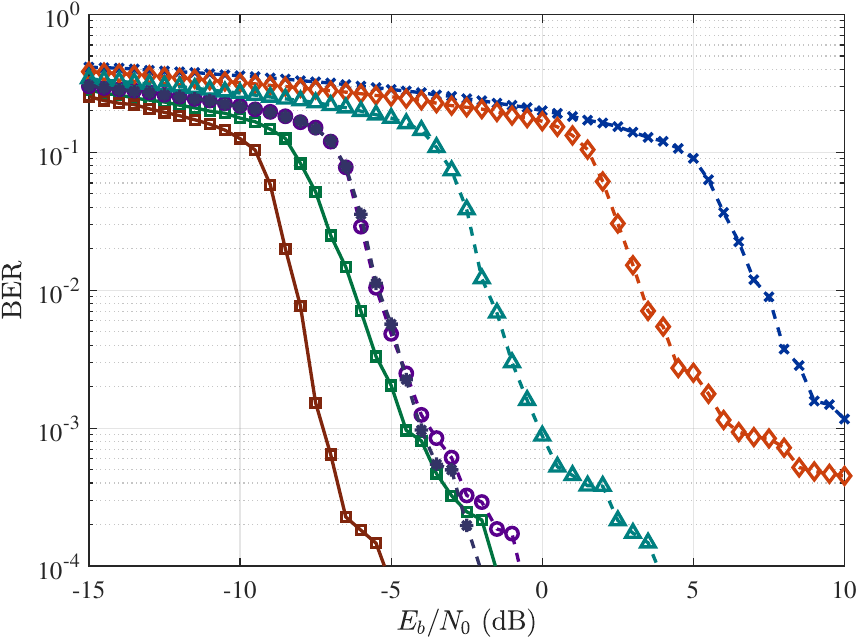}
         \caption{BER performance.}   
         \label{Fig: ber_PAPR_constrained}
     \end{subfigure}
     \begin{subfigure}[b]{0.495\linewidth}
         \centering
         \includegraphics[width=0.9\textwidth]{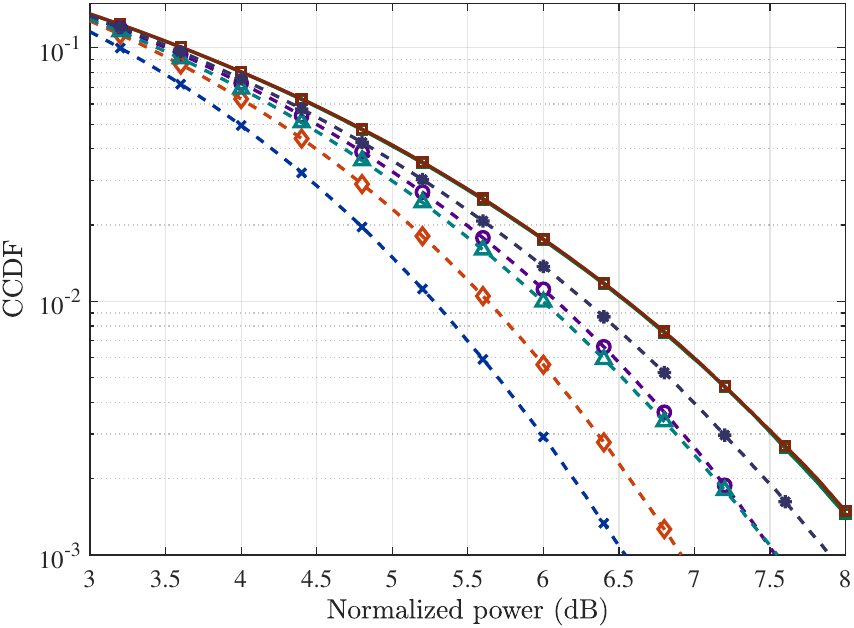}
         \caption{CCDF of the normalized power.}
    \label{Fig: ccdf_power}
     \end{subfigure}

\caption{BER performance and CCDF of the normalized power under a pilot-free and CP-free configuration in the CDL-C channel model at $120$ km/h.}
\label{Fig: performance_diff_papr_constrain}
\end{figure*}

In the previous subsection, to simplify the performance comparison, the PAPR constraint is not considered. In this subsection, we conduct an in-depth evaluation of system performance under the PAPR constraint. The complementary cumulative distribution function (CCDF) of the normalized power samples is used to characterize the power behavior. As demonstrated in~\cite{aoudia2021end}, the geometric shaping exhibits a nearly identical PAPR distribution to conventional QAM modulation. Therefore, this work focuses on evaluating the PAPR reduction of the time-domain oversampled signal in a pilot-free and CP-free system.

To provide a more intuitive illustration of constellation point distributions under different PAPR constraints, kernel density estimation (KDE) is applied, as shown in Fig.~\ref{Fig: papr_kde}. As the PAPR constraint becomes more stringent, the amplitude of most constellation points except those serving as anchors gradually decreases, leading to a denser concentration of points around the origin. However, this optimization is accompanied by a degradation in BER performance due to reduced signal detection accuracy. Fig.~\ref{Fig: performance_diff_papr_constrain} presents a comparison of the BER and the CCDF curves corresponding to different PAPR constraint settings with a modulation order of $M = 6$. The solid lines represent networks trained without PAPR constraints, while the dashed lines correspond to networks trained with PAPR constraints. The BER is considerably high at $\epsilon_P = 6.5\text{ dB}$. By relaxing the PAPR constraint to $\epsilon_P = 8.0\text{ dB}$, the constellation points near the origin become more dispersed, leading to further BER improvement. The resulting performance approaches that of the E2E system without PAPR constraints~\cite{aoudia2021end}, and also outperforms the conventional transceiver shown in Fig.~\ref{fig:cdlc_ber_throughput}(\subref{fig:ber_120}). In addition, we simulate a hybrid scheme that combines the conventional clipping technique with the PAPR-constrained training, where the clipping rate is set to 1. The results show that the hybrid approach further reduces PAPR with only a slight BER degradation at $\epsilon_P = 8.0\text{ dB}$. This suggests that combining conventional PAPR reduction methods with learning-based optimization is a promising direction. When the PAPR constraint is not considered and the training is conducted using the CE loss only, the learned constellation achieves a PAPR of approximately $8.25\text{ dB}$ at the $10^{-3}$ CCDF level while providing the highest BER gain.

\section{Conclusion and Future Work}
In this paper, we have proposed an adaptive E2E transceiver architecture tailored for pilot-free and CP-free OFDM systems. By incorporating AI-enabled geometric constellation shaping and a neural receiver, the framework significantly reduces BER while achieving a $26.4\%$ improvement in throughput over conventional systems. A lightweight plug-and-play channel adapter further enhances adaptability under dynamic channels, achieving comparable BER performance to full fine-tuning while updating only $3.5\%$ of the parameters. Furthermore, a modulation-order scalable strategy is proposed, enabling a unified model to support multiple modulation orders within a single architecture, which reduces model storage overhead by up to $75\%$. To address the PAPR challenge in OFDM systems, constrained E2E training is employed, ensuring compliance with PAPR limits without introducing additional bandwidth overhead.
Extensive simulations across 3GPP-compliant channel models and mobility scenarios validate the proposed transceiver’s superior performance in BER, throughput, online adaptability, storage overhead, and PAPR reduction. These results highlight its potential for AI-native air interface design in NextG systems, promoting the feasibility of pilot-free and CP-free transmission for standardization and practical deployment. 

Future research may focus on the integration of efficient intelligent channel coding and decoding techniques, lightweight model design~\cite{zhang2025ghost}, and the extension to emerging channel models and multi-user MIMO systems~\cite{3gpp_apple,you2025next}. 
In particular, in the MU-MIMO scenario, we can exploit multi-user time–frequency resource multiplexing, where the resource grids originally reserved for pilots are instead occupied by the designed multi-user data symbols to further enhance spectral utilization. Each user is further equipped with a trainable constellation geometry and bit labeling strategy, enabling adaptive symbol mapping and improved transmission efficiency.

\bibliographystyle{IEEEtran}
\bibliography{myref}

\vfill

\end{document}